\documentclass[pre,onecolumn,superscriptaddress,aps,10pt]{revtex4-2}

\usepackage[T1]{fontenc}
\usepackage{amsfonts}
\usepackage{amsmath}
\usepackage{amssymb}
\usepackage{graphicx}
\usepackage{dcolumn}
\usepackage{times}
\usepackage{color}
\usepackage{bbm}
\usepackage{comment}

\DeclareSymbolFont{matha}{OML}{txmi}{m}{it}
\DeclareMathSymbol{\vi}{\mathord}{matha}{118}

\newcommand{\e}{\varepsilon}
\newcommand{\beq}{\begin{equation}}
\newcommand{\eeq}{\end{equation}}

\newcommand{\relerr}{\epsilon}

\newcommand{\lam}[1]{\lambda^{(#1)}}

\begin{document}

\title{Series approximation of discrete-soliton solutions of the DNLS equation}

\author{V.\ Tsiantou}
\affiliation{Department of Mathematics, University of the Aegean, Karlovasi, 83200 Samos, Greece}
\author{V.\ Koukouloyannis}
\affiliation{Department of Mathematics, University of the Aegean, Karlovasi, 83200 Samos, Greece}
\author{K.\ D.\ Housiadas}
\affiliation{Department of Mathematics, University of the Aegean, Karlovasi, 83200 Samos, Greece}
\author{P.G.\ Kevrekidis}
\affiliation{Department of Mathematics and Statistics, University of Massachusetts, Amherst MA 1003-4515, USA}

\begin{abstract}
We study standing waves of the discrete nonlinear Schr\"{o}dinger (DNLS) equation
by means of perturbative series expansions in the inter-site coupling parameter $\varepsilon$.
Power-series approximations are derived for three representative configurations: a single-site
soliton and the two two-site ones, namely the in-phase and the out-of-phase configurations. 
Although higher-order truncations are increasingly accurate for small $\varepsilon$, the expansions
possess a small radius of convergence and cease to be useful well before $\varepsilon={\cal O}(1)$.
To extend their range of validity we apply the Euler convergence-acceleration method, constructing
an optimally transformed series for each lattice site individually. The transformed solution is shown to have substantially better convergence radius and preserves the localization of the solution much better than the original series solution. 
We then insert  the series profiles  into the linearization matrix of the DNLS equation in order to obtain improved
approximations of the stability eigenvalues, with particular attention to the unstable real
eigenvalue of the in-phase configuration. Optimal truncation of the corresponding expansion yields
a closed-form ${\cal O}(\e^{5/2})$ estimate, which significantly extends the known ${\cal O}(\e^{1/2})$ leading-order result to considerably larger couplings, while the
semi-numerical use of the accelerated profiles reproduces the numerically computed eigenvalue to values of $\e$ of ${\cal O}(1)$. Throughout, both the solution profiles and
their stability eigenvalues are benchmarked against the numerical solution obtained through a
continuation procedure from the anti-continuum limit.
\end{abstract}

\maketitle

\section{Introduction - The model}
Coherent structures arise ubiquitously in nonlinear dispersive media, where the interplay between
dispersion (or discrete diffraction) and nonlinearity supports robust, spatially localized excitations that
propagate or remain pinned without dispersing~\cite{ablowitz2011nonlinear}. Such structures --- solitons, vortices, kinks and
breathers --- have been investigated extensively over the past several decades in both continuous
and discrete settings, with applications ranging from nonlinear optics and Bose--Einstein
condensates (BECs) to biophysics, condensed matter and electrical
lattices~\cite{ablowitz2011nonlinear,Scott2003, AblowitzPrinariTrubatch, Dauxois2006,Kevrekidis2009,KevrekidisBEC2015}. Discreteness introduces
qualitatively new phenomena absent in the continuum limit, most notably the existence of
exponentially localized, time-periodic solutions known as discrete breathers~\cite{MacKayAubry1994} or, in the Hamiltonian
gauge of the discrete nonlinear Schr\"{o}dinger (DNLS) equation, 
discrete solitons in the form of standing waves~\cite{Kevrekidis2009}.
It is relevant to point out that such  discrete solitonic structures have been extensively
studied for multiple decades as evidenced by corresponding reviews~\cite{LEDERER20081,morsch}, they continue to be of current interest to this day.
This is, to a significant degree, due to the fact that novel settings manifesting
their robust existence continue to emerge, ranging from accordion BEC
lattices~\cite{haller}, to optically induced lattices in atomic vapors~\cite{Vulic2026}.

The rigorous foundation of the theory of discrete breathers was laid by MacKay and
Aubry~\cite{MacKayAubry1994}, who proved that, starting from the so-called anti-continuum (AC) ---also referred to, early on, as
anti-integrable--- limit of uncoupled nonlinear oscillators, time-periodic and exponentially localized
solutions persist for small values of the coupling parameter under generic non-resonance conditions.
This anti-continuum-limit construction has since become a canonical framework for proving the
existence of multi-site breathers and discrete solitons in a variety of lattice models, and
extensive reviews of the relevant phenomenology can be found in
Refs.~\cite{FlachWillis1998, FlachGorbach2008, Aubry1997, Aubry2006}. In the case of the DNLS
equation, which is the focus of the present work, the anti-continuum limit corresponds to vanishing
inter-site coupling $\varepsilon \rightarrow 0$, where the trivial solutions consist of arbitrary
superpositions of (without loss of generality) unit-amplitude excitations and empty sites; these solutions can be continued to
$\varepsilon > 0$ to yield families of discrete solitons parametrized by $\varepsilon$~\cite{Kevrekidis2009}.

Beyond the question of existence, the linear stability of these structures has attracted
considerable attention, since stability ultimately determines which configurations are dynamically robust and, hence, the best candidates toward being physically
observable. For the one-dimensional DNLS lattice, a systematic stability theory of discrete solitons bifurcating
from the anti-continuum limit was developed 
in the work of~\cite{PelinovskyKevrekidisFrantzeskakis2005a}, who classified the unstable eigenvalues
of arbitrary in-phase/anti-phase multi-site configurations and proved that, near $\varepsilon = 0$,
the only linearly stable configurations are those formed by alternating anti-phase excited nodes.
Companion results for discrete vortices in two-dimensional DNLS lattices were obtained in
Ref.~\cite{PelinovskyKevrekidisFrantzeskakis2005b} with generalizations
in three-dimensional settings in~\cite{LUKAS2008339}, and the associated internal modes were analysed
in Ref.~\cite{PelinovskySakovich2011}. Arguably, 
the most elemental multi-site configuration is that with two excited nodes,
which is unstable when they bear the same phase (phase-difference of $0$
or in-phase configuration) and spectrally stable near the anti-continuum
limit when they bear a phase-difference of $\pi$ (out-of-phase configuration).
The unstable real eigenvalue associated with
an in-phase two-site soliton was shown to scale as 
$\lambda = 2\sqrt{\varepsilon}$ to leading order~\cite{PelinovskyKevrekidisFrantzeskakis2005a},
a result that we will use as a benchmark below. 

It is worth noting in passing a sequence of parallel developments in the
realm of genuine discrete breathers (involving infinitely many frequencies
via harmonic generation) in Klein-Gordon and Fermi-Pasta-Ulam-Tsingou lattices~\cite{FlachGorbach2008}.
In particular, following on the seminal earlier work of~\cite{ARCHILLA2003235}
and of~\cite{Aubry2006}, an alternative approach to the stability of
multibreathers in Klein--Gordon chains has been developed in  ~\cite{KoukouloyannisKevrekidis2009}. It has been based on the averaging theory for Hamiltonian systems in
action--angle variables originally introduced by MacKay and coworkers~\cite{AhnMacKaySepulchre2001,Koukouloyannis20051021}. 
This provides an ${\cal O}(\sqrt{\varepsilon})$ estimate of the characteristic Floquet exponents and
yields, for soft (hard) on-site nonlinearities, simple criteria determining which sign of the phase
difference between adjacent excited sites supports stable multibreather configurations. The
equivalence between these two complementary approaches was subsequently established in
Ref.~\cite{Cuevas2011Multibreather}, and the framework has been extended in several directions,
including interactions beyond nearest neighbors~\cite{KoukouloyannisKevrekidisCuevasRothos2013} and
resonances and bifurcations of breathers with holes~\cite{PelinovskySakovich2012}.

A common feature of all the analytical approaches in the vicinity of the
AC limit is that they rely on a perturbative
expansion of the localized solution, and of its stability eigenvalues, around  
$\varepsilon = 0$. Such expansions capture the solution accurately for sufficiently small
$\varepsilon$, but they lose their predictive power well before $\varepsilon$ reaches
${\cal O}(1)$, where numerical continuation typically remains perfectly feasible. Direct
approximations of DNLS solitons that go beyond the leading-order anti-continuum construction have
also been proposed~\cite{CuevasJamesKevrekidisMalomedSanchezRey2008}. This raises a natural
question: to what extent can one extend the predictive range of the perturbative series, both for
the soliton profile itself and for its stability eigenvalues, by means of convergence-acceleration
techniques applied to the truncated expansion? Series-acceleration methods --- Euler, Borel,
Pad\'{e}, Shanks --- have a long history in applied mathematics and asymptotic
analysis~\cite{BenderOrszag1999, Hardy1949, Borel1899, Pade1892, Shanks1955, Wynn1956,
VanWijngaarden1965, Levin1973, Weniger1989, BakerGravesMorris1996, BrezinskiRedivoZaglia1991,
Sidi2003}, and have been shown to improve dramatically the accuracy of perturbative expansions in a
variety of physical contexts; a recent example in the context of viscoelastic flows can be found in
Ref.~\cite{Housiadas20171}.

In the present work we pursue exactly this program for the 
prototypical example of the cubic DNLS equation. We compute
high-order perturbative series for three of the most elemental and 
representative configurations: a single-site discrete
soliton, an in-phase two-site soliton and an out-of-phase two-site soliton. 
We use the Euler
transform to extend the validity range of these expansions. We then feed the accelerated convergence
soliton
profiles into the linearization (stability) matrix~\cite{Kevrekidis2009} in order to obtain improved
estimates of the unstable eigenvalues, which we compare against full numerical continuation of both
the soliton profile and its spectrum. As we shall see, this procedure substantially enlarges the
parameter range in which the perturbative series faithfully reproduces the numerical result, and it
also delivers a compact analytical formula for the  stability eigenvalues of the two-site configurations
that improves on the known leading-order estimates.

This paper is organized as follows. In Sec.~\ref{sec:Euler} we briefly present the Euler
transformation algorithm, in both its standard and its ``improved'' form. In Sec.~\ref{sec:solutions}
we derive series approximations for the three discrete-soliton configurations considered and assess
their accuracy against numerical continuation. In Sec.~\ref{sec:stability} we use these
approximations to compute the stability eigenvalues of the two-site configurations, and we derive improved analytical estimates for them. Finally,
Sec.~\ref{sec:conclusions} summarizes our conclusions and outlines directions for future work.

\section{The Euler transform}\label{sec:Euler}
The Euler transformation is a linear method for accelerating the convergence of a (sign-)alternating
series. Consider the series
$$ {\cal S}= \sum_{j=0}^{\infty}(-1)^{j}\alpha _{j}=\alpha_{0}-\alpha_{1}+\alpha_{2}-\alpha_{3}+\ldots ,\quad \alpha_{j}>0. $$
Then, the Euler transformation is defined by the series~\cite{Euler_initial}
\begin{equation}
{\cal E}=\sum_{j=0}^{\infty}\dfrac{(-1)^{j}\Delta^{j}\alpha_{0}}{2^{j+1}},
\label{eq:euler}
\end{equation}
where $ \Delta^{j}\alpha_{m}$ denotes the forward difference
$ \Delta^{j}\alpha_{m}=\sum_{k=0}^{j}(-1)^{k}\left( \begin{array}{c} j \\ k \end{array} \right)\alpha_{m+j-k} $,
and $\left( \begin{array}{c} j \\ k \end{array} \right)$ is the binomial coefficient. As has been shown (see e.g. ~\cite{Hardy1949} \S8), 
this series can have better convergence properties than ${\cal S}$. In our work we are going
to apply the transform to truncated series of $n$ terms, denoted by ${\cal S}_n$ and ${\cal E}_n$
respectively.

{It is relevant to note that the Euler series can be rewritten as:
\begin{eqnarray}
{\cal E}
=
\frac{\alpha_0}{2}
+\frac{\alpha_0-\alpha_1}{4}
+\frac{\alpha_0-2\alpha_1+\alpha_2}{8}
+\cdots.
\label{Eul1}
\end{eqnarray}
The point of the improvement that this reshuffling induces is that for a reasonably smooth, slowly decreasing sequence $\alpha_j$, the finite differences
$\alpha_0-\alpha_1,\quad
\alpha_0-2\alpha_1+\alpha_2,\quad\ldots$
typically become small, while you also acquire the geometric factors $2^{-j-1}$. That is where the acceleration in the series convergence comes from.}

In~\cite{Euler_transform} a more practical approach to derive the transformation~\eqref{eq:euler} is
presented. In order
to demonstrate this approach, we initially consider a truncated series with five terms,
$\displaystyle {\cal S}_5=\sum_{j=0}^4(-1)^j\alpha_j$, and construct the following triangular
table:
\begin{center}
\begin{tabular} {clc}
$\begin{array}{ccccc}
A_{1} &  &  &  &  \\
 & B_{1} & &  &    \\
A_{2} & &C_{1}  &  &  \\
 & B_{2} &  & D_{1} &  \\
A_{3} & &C_{2} & &E_{1}\\
 & B_{3} &  &D_{2}  & \\
A_{4} &  &C_{3}  &  & \\
 &B_{4}  &  &  & \\
A_{5}& & & & \\
\end{array}$
&,\qquad where\qquad $ $&
$\begin{array}{ccc}
A_{1} & = & \alpha_{0}  \\
A_{2} & = & -\alpha_{1} \\
A_{3} & = & \alpha_{2} \\
A_{4} & = & -\alpha_{3} \\
A_{5} & = & \alpha_{4}\\
\end{array}.$
\end{tabular}
\end{center}
The elements of the first column are the successive signed terms of ${\cal S}_5$, while the entries
of the following columns $B_{i}, C_{i},\ldots$ are the mean values of the two preceding elements,
i.e.\ $B_{i}=(A_{i+1}+A_{i})/2$, $C_{i}=(B_{i+1}+B_{i})/2$, and so on. We refer to the elements
carrying the same index as a ``primary diagonal'' of the table, e.g.\ $(A_{1}, B_{1}, C_{1}, D_{1},
E_{1})$, while the ``mirroring'' elements with increasing indices form a ``secondary diagonal'',
e.g.\ $(E_1, D_2, C_3, B_4, A_5)$.

The truncated Euler-transformed series ${\cal E}_5$ is obtained by summing the elements of the first
primary diagonal and halving the result,
\beq{\cal E}_5= \frac{1}{2}\left(A_{1}+B_{1}+C_{1}+D_{1}+E_{1}\right)=\frac{1}{32}\left({31 \alpha _0}-{26 \alpha _1}+{16 \alpha _2}-{6 \alpha _3}+{\alpha_4}\right).\label{eq:e5}\eeq
The right side of (\ref{eq:e5}) can be extracted either by the algorithm described above or directly by applying \eqref{eq:euler}. 

However, according to~\cite{Euler_transform}, a more accurate
approximating series can be obtained using the same construction. The primary
diagonals are examined in sequence, and the first one whose elements are all smaller in magnitude
than the corresponding elements of the next secondary diagonal is designated as the ``principal
diagonal'' of the table. The best-converging approximation is then obtained by summing the elements
of the $A$ column down to the principal diagonal and adding half the sum of the elements of the
principal diagonal itself. We denote this transform by ${\cal E}_{5i}^m$, where the subscript
``$i$'' stands for ``improved'' and the superscript ``$m$'' labels the principal diagonal.
For example, if $\vert C_1\vert>\vert B_2\vert$, the first primary diagonal is not the principal one
and the second one is examined; if all of its elements are smaller in magnitude than the
corresponding elements of the third primary diagonal, i.e.\ $|B_2|<|A_3|$,
$|C_2|<|B_3|$ and $|D_2|<|C_3|$, then it is the principal
diagonal and the fastest-converging approximation is
${\cal E}_{5i}^{2}=A_{1}+\frac{1}{2}\left( A_{2}+B_{2}+C_{2}+D_{2}\right)$.
Five different approximating series can therefore be constructed for $n=5$; the corresponding
conditions and the associated approximations are summarized in Table~\ref{tab:Eulers}. Note that the
improved transform ${\cal E}_{5i}^1$ coincides with the standard Euler transform ${\cal E}_5$,
while ${\cal E}_{5i}^5$ coincides with the original truncated series ${\cal S}_5$.

\begin{table}[h]
\centering
\renewcommand{\arraystretch}{1.5}
\begin{tabular}{cccl}
\hline
&Condition& &$ $\qquad\quad Best approximating series\\
\hline
if&$  |B_{1}| <   |A_{2}| $ and $  |C_{1}| <   |B_{2}| $ and $  |D_{1}| <   |C_{2}| $ and $  |E_{1}| <   |D_{2}| $&
$\rightarrow$&
\ \ ${\cal E}_{5i}^{1}={\cal E}_{5}= \frac{1}{2}\left(A_{1}+B_{1}+C_{1}+D_{1}+E_{1}\right) $\\
\hline
else if&$  |B_{2}| <   |A_{3}| $ and $  |C_{2}| <   |B_{3}| $ and $  |D_{2}| <   |C_{3}| $ &
$\rightarrow$&
\ \ ${\cal E}_{5i}^{2}=A_{1}+\frac{1}{2}\left( A_{2}+B_{2}+C_{2}+D_{2}\right) $\\
\hline
else if&$  |B_{3}| <   |A_{4}| $ and $  |C_{3}| <   |B_{4}| $
&
$\rightarrow$&
\ \ ${\cal E}_{5i}^{3}=A_{1}+A_{2}+\frac{1}{2}\left( A_{3}+B_{3}+C_{3}\right) $\\
\hline
else if&$  |B_{4}| <   |A_{5}| $&
$\rightarrow$&
\ \ ${\cal E}_{5i}^{4}=A_{1}+A_{2}+A_{3}+\frac{1}{2}\left( A_{4}+B_{4}\right) $\\
\hline
&else &$\rightarrow$&
\ \ ${\cal E}_{5i}^{5}={\cal S}_{5}=A_{1}+A_{2}+A_{3}+A_{4}+A_{5}$\\
\hline
\end{tabular}
\caption{Selection rule for the improved Euler transformation, illustrated for a five-term
truncation. The conditions are tested in the order shown, and the first one that is satisfied fixes
the principal diagonal and hence the approximating series.}
\label{tab:Eulers}
\end{table}
Note that, in some works, just one of the ${\cal E}_{5i}^m$ transforms is used, to avoid the complexity of the selection rule. 
For example in \cite{Housiadas20171} the ${\cal E}_{5i}^2$ is used, since it provides a good balance between accuracy and computational complexity for the problems considered in that work. 

\section{Application to discrete-soliton solutions}\label{sec:solutions}
\subsection{Mathematical Setup}
Our paradigmatic setting involves
the one-dimensional cubic DNLS equation in the form
\begin{equation}
\mathbbm{i} \dot{u}_{i}+\e\left( u_{i+1}-2u_{i}+u_{i-1}\right)  + \vert u_{i}\vert ^{2}u_{i}=0,\quad i\in\mathbb{Z},
\label{dnls}
\end{equation}
where $\mathbbm{i}$ denotes the imaginary unit, $i$ the lattice site and $\e$ the inter-site
coupling. We look for stationary solutions of the form $u_{i}(t)=\vi_{i}e^{\mathbbm{i}\omega t}$,
where $\vi_{i}$ is time independent. Substituting this ansatz into~\eqref{dnls} and applying the
scaling transformation $\vi_{i}\mapsto\sqrt{\omega}\,\vi_{i}$, $\e\mapsto\e/\omega$, we obtain the
monoparametric algebraic form of the equation,
\begin{equation}
-\vi_{i}+\varepsilon \left( \vi_{i+1}-2\vi_{i}+\vi_{i-1}\right)  +\vert \vi_{i}\vert ^{2}\vi_{i}=0,
\label{algebraic}
\end{equation}
which is the form considered throughout this work.

In order to approximate specific configurations of discrete-soliton solutions of~\eqref{algebraic},
we consider series expansions of the solution of the form
\beq \vi_{i}=\vi_{i}^{(0)}+\varepsilon \vi_{i}^{(1)}+\varepsilon ^{2}\vi_{i}^{(2)}+\ldots+\varepsilon ^{n}\vi_{i}^{(n)},
\label{exp0}\eeq
where $i$ denotes the lattice site and $n$ the order of the approximation. Inserting this truncated
series, for all the sites considered, into~\eqref{algebraic} yields the truncated-series
approximation of the solution for the whole lattice. The aim of this paper is to compute such series
approximations for three specific discrete-soliton configurations: the single-site (on-site)
configuration and the two two-site ones, known as in-phase and out-of-phase. Since the resulting
series are only useful over a small range of $\e$, we employ convergence-acceleration techniques in
order to improve the accuracy of the resulting approximations. Specifically, we use the Euler
transform, a linear transformation applicable to sign-alternating series. Having improved the
accuracy of the computed profiles, we then use them to sharpen the approximation of the
corresponding stability eigenvalues obtained in
Ref.~\cite{PelinovskyKevrekidisFrantzeskakis2005a}.

\subsection{Single-site configuration}
In order to obtain an approximation of the single-site discrete-soliton solution we consider the
following truncated series, one for each site of the lattice,
\beq
\begin{array}{rcl}
\vi_{0}&\simeq&1 +\varepsilon \vi_{0}^{(1)}+\varepsilon ^{2}\vi_{0}^{(2)}+\ldots+\varepsilon ^{n}\vi_{0}^{(n)},\\[0.1cm]
\vi_{i}&\simeq&0+\varepsilon \vi_{i}^{(1)}+\varepsilon ^{2}\vi_{i}^{(2)}+\ldots+\varepsilon ^{n}\vi_{i}^{(n)},\quad \forall i\neq0.\label{exp1}
\end{array}\eeq
This choice guarantees that in the AC limit $\e\rightarrow0$ we recover the trivially
localized solution $\{\ldots,0,0,1,0,0,\ldots\}$. Substituting an approximation of degree $n$
into Eq.~\eqref{algebraic} and applying perturbation theory order by order, we obtain the corresponding
truncated-series solution of Eq.~\eqref{algebraic} in powers of $\varepsilon$. We denote the set of
these solutions for all the sites of the lattice by $S_n$. For $n=5$ we obtain
$$S_5=\{\ldots, 0, \vi_{5}, \vi_{4}, \vi_{3}, \vi_{2}, \vi_{1}, \vi_{0}, \vi_{1}, \vi_{2}, \vi_{3}, \vi_{4}, \vi_{5}, 0, \ldots\},$$
with
\beq\begin{array}{rclcrcl}
\vi_{0}&\simeq&1+\varepsilon-\dfrac{3\varepsilon ^{2}}{2}+\dfrac{7\varepsilon ^{3}}{2}-\dfrac{85\varepsilon ^{4}}{8}+\dfrac{287\varepsilon^{5}}{8}  &$ $\qquad$ $ &\vi_{3}&\simeq&\varepsilon ^{3}-5\varepsilon^{4}+\dfrac{41\varepsilon^{5}}{2},\\[0.2cm]
\vi_{1}&\simeq&\varepsilon-\varepsilon ^{2}+\dfrac{5\varepsilon ^{3}}{2}-\dfrac{15\varepsilon^{4}}{2}+\dfrac{195\varepsilon^{5}}{8} & &
\vi_{4}&\simeq &\varepsilon^{4}-7\varepsilon^{5},\\[0.2cm]
\vi_{2}&\simeq &\varepsilon ^{2}-3\varepsilon ^{3}+\dfrac{19\varepsilon^{4}}{2}-\dfrac{63\varepsilon^{5}}{2}& &
\vi_{5}&\simeq &\varepsilon^{5}.
\end{array}\label{eq:S_5_single}\eeq
The choice $n=5$ is dictated by the localization of the solution, which ensures that a few sites
around the central one suffice to approximately yet quite accurately 
describe the discrete soliton. Indeed, the expansion of each site
starts one order higher than that of its inner neighbour, which is a direct consequence of the
exponential localization of the solution; consequently, the higher the degree of the expansion, the
more non-zero sites are generated. In the present case we obtain five non-zero sites on each side of
the central one, which is sufficient for an adequate representation of the true solution.

In order to assess the accuracy of these approximations we perform a numerical continuation of the
trivially localized configuration with respect to $\e$. We treat this numerical solution as the ground
truth and denote it by $N$. The step used in the parametric 
continuation procedure is $\Delta\e=0.001$ and the solution was approximated to machine precision. In what follows, the total range of $\e$ considered varies depending on the aspect of the problem we examine each time.
\begin{figure}[!b]
\includegraphics[width=0.32\textwidth]{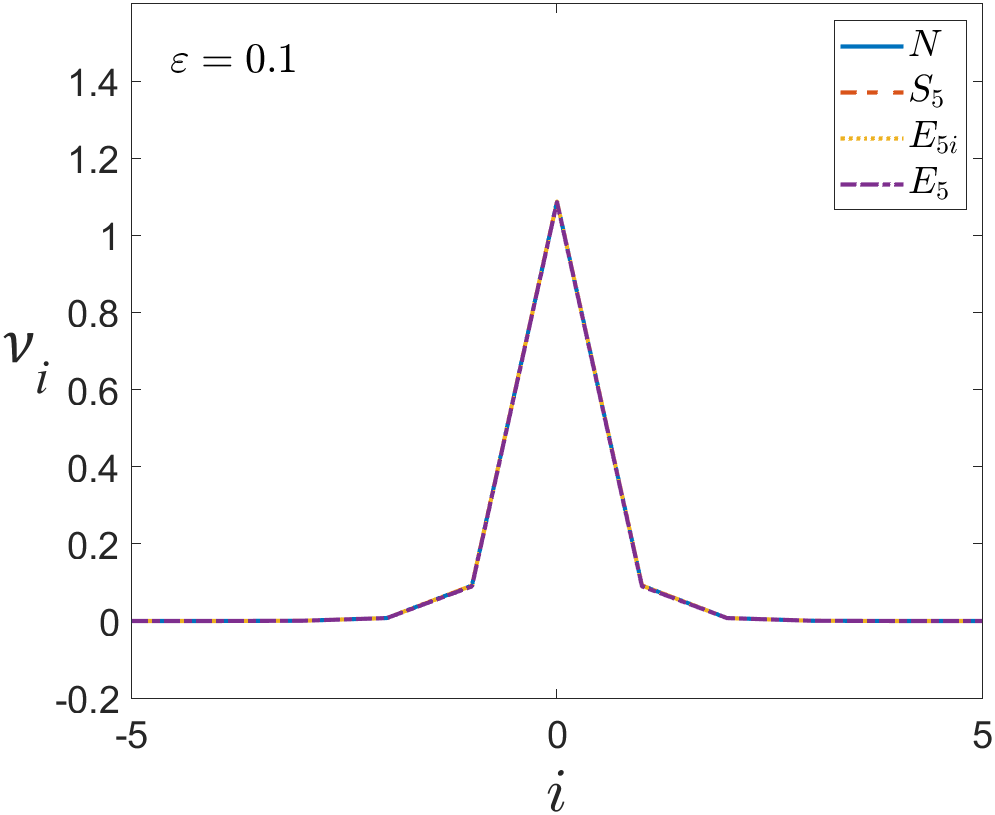}
\includegraphics[width=0.32\textwidth]{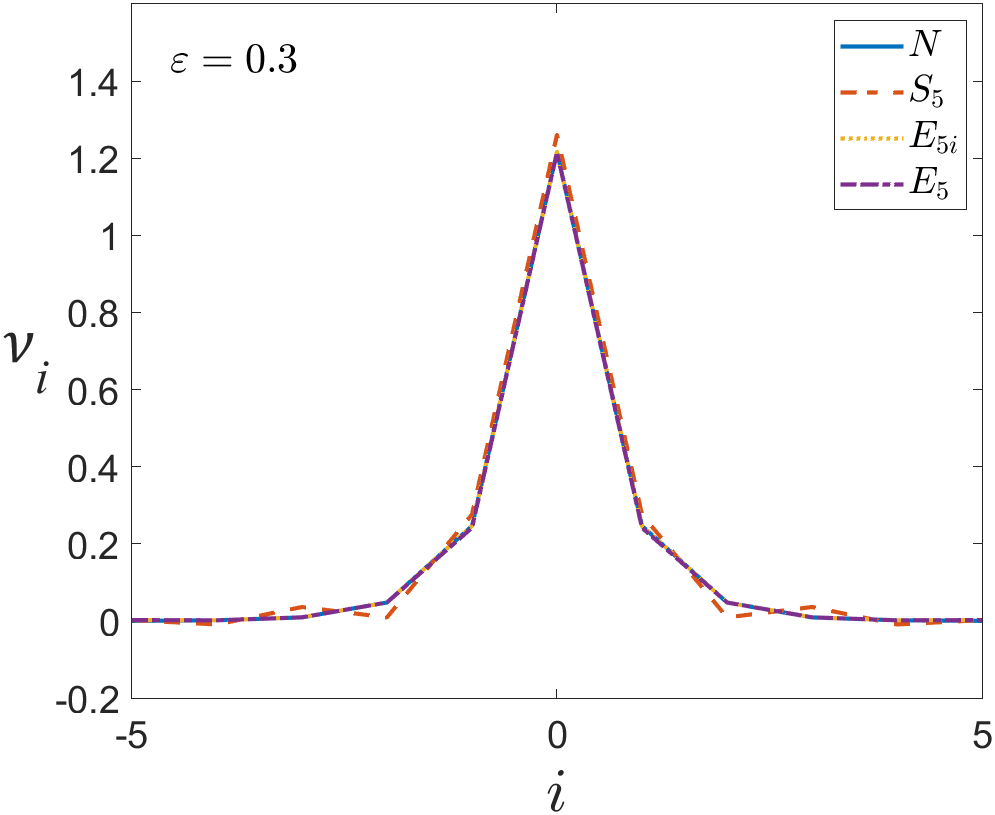}
\includegraphics[width=0.32\textwidth]{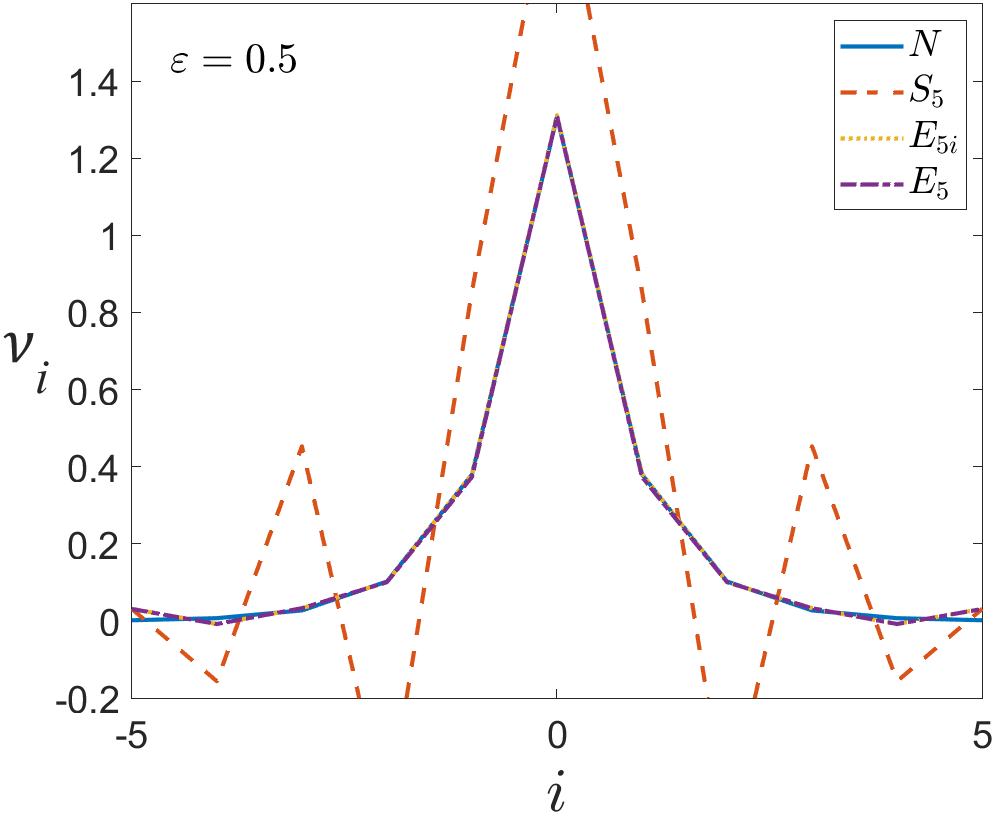}
\caption{Profile of the single-site discrete-soliton solution on a linear scale, for three characteristic values of
the coupling parameter, namely $\e=0.1, 0.3, 0.5$. The numerical solution N (blue, solid) is compared with the standard $n=5$
term series expansion $S_5$ (red, dashed), the Euler transform result ${E}_5$ (purple, dash-dotted) and
the improved variant the latter ${E}_{5i}$ (yellow, dotted).
}
\label{fig:portrait_single}
\end{figure}

Since all the perturbation expansions derived below are sign-alternating, the Euler transformation
provides a natural linear acceleration procedure. In what follows we apply both the standard and
the improved Euler transformations. Note that the perturbative solution provides a different number
of terms for each site, so that transforms of different length are required
(${\cal E}_5, {\cal E}_4, {\cal E}_3, \ldots$). Although the expansion of $\vi_0$ contains six
terms, the Euler transform is applied to the $\vi_0-1$ series, i.e., to the terms of ${\cal O} (\e)$ and higher orders; in this way, the sequence fed to the transform is strictly alternating and
effectively has five terms. Then it is added to the constant term $1$ to get the transformed expression for $\vi_0$. We denote the set of all these transformed solutions by $E_5$, and the
corresponding set for the improved transformation by $E_{5i}$; 
for a general degree $n$ of
approximation the analogous sets are denoted by $E_n$ and $E_{ni}$. The transformed series obtained
for each site and for each range of $\e$ are listed in Appendix~\ref{EulerSingle}.

Figure~\ref{fig:portrait_single} shows the profile of the numerically computed discrete soliton $N$
together with the $S_5$, $E_5$ and $E_{5i}$ approximations. For low coupling ($\e=0.1$) all the
series provide a reasonable approximation. For $\e=0.3$, the 
$S_5$ solution starts to deviate from $N$ and
its sign-alternating character becomes visible. Finally, for $\e=0.5$ the $S_5$ solution no longer accurately
represents the true solution, while only a small deviation at the edges of the lattice starts to appear
in the $E_5$ and $E_{5i}$ approximations as well.

\begin{figure}
\includegraphics[width=0.32\textwidth]{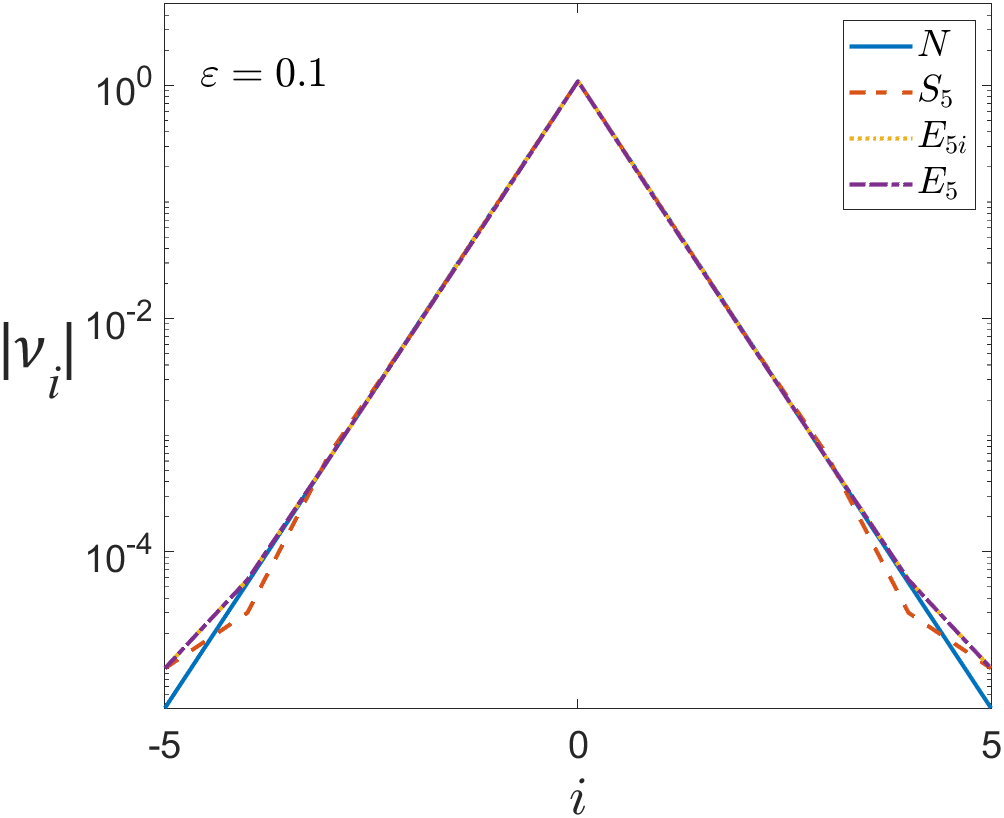}
\includegraphics[width=0.32\textwidth]{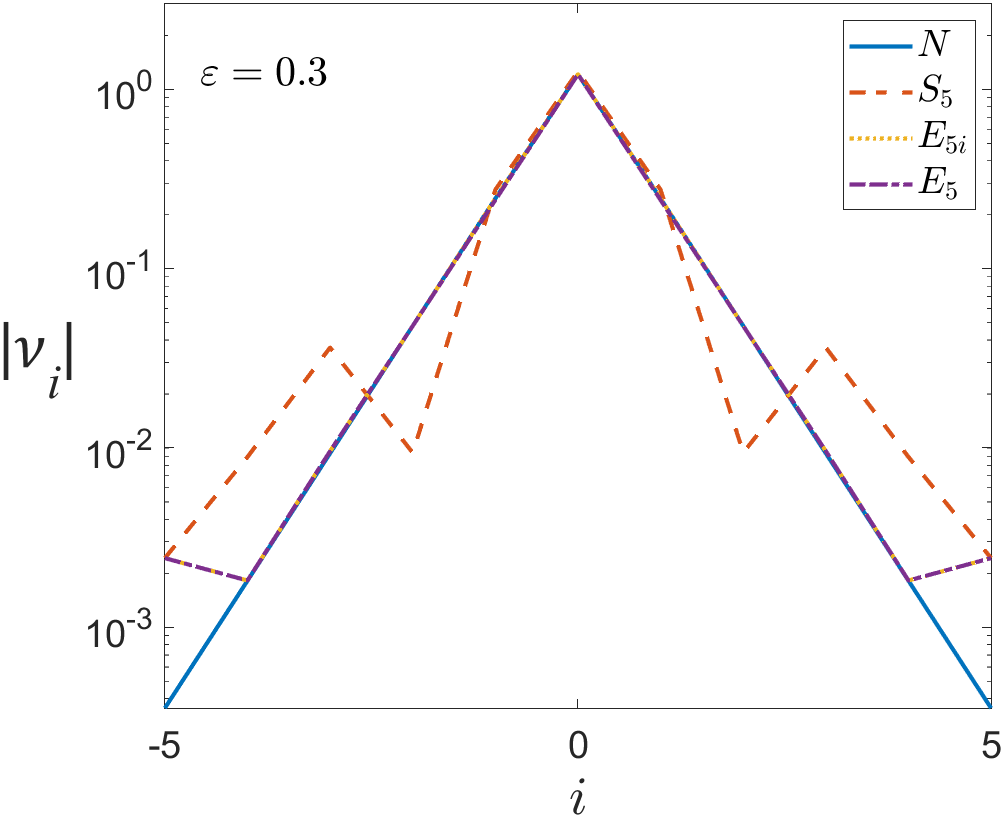}
\includegraphics[width=0.32\textwidth]{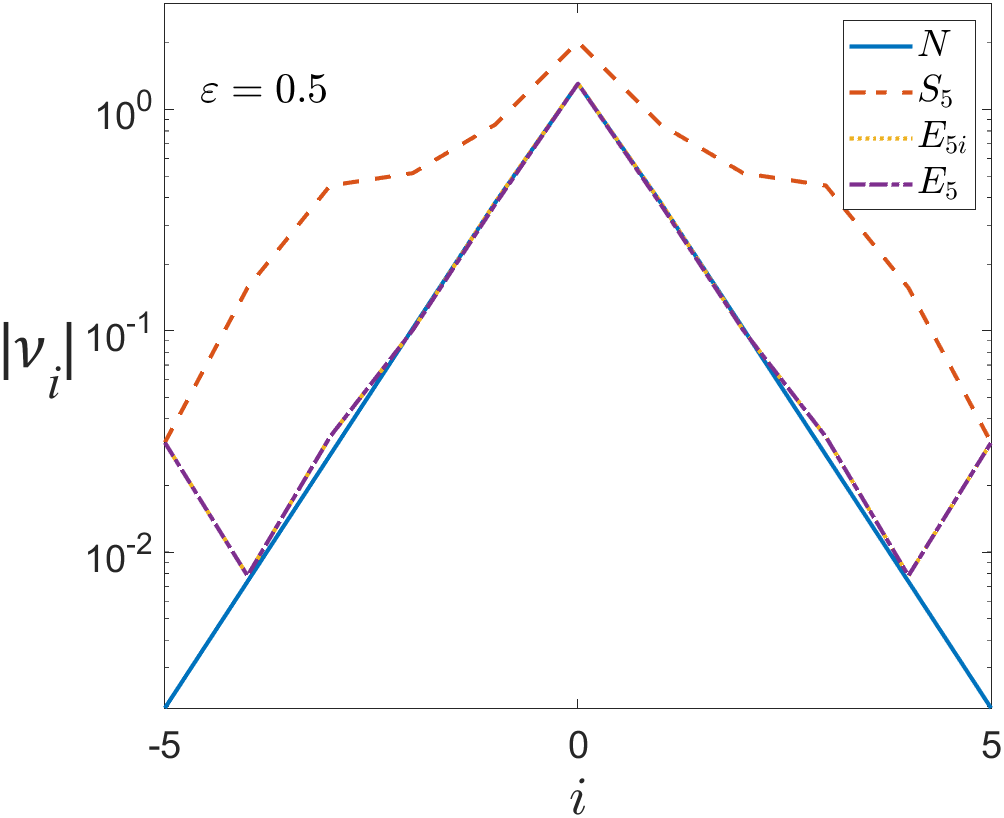}
\caption{Profile of the single-site discrete-soliton solution on a semilogarithmic scale, for three characteristic values of
the coupling parameter, namely $\e=0.1, 0.3, 0.5$. The numerical solution N (blue, solid) is compared with the standard $n=5$
term series expansion $S_5$ (red, dashed), the Euler transform result ${\cal E}_5$ (purple, dash-dotted) and
the improved variant the latter ${\cal E}_{5i}$ (yellow, dotted).}
\label{fig:portrait_single_log}
\end{figure}

In order to resolve these deviations more clearly we replot the same profiles on a semilogarithmic
scale in Fig.~\ref{fig:portrait_single_log}. The exponential localization of the solution $N$ is
evident in all three panels. The deviations at the edges of the lattice already appear at
$\e=0.1$: the $S_5$ expansion of $\vi_5$ consists of a single term and, as
can be seen in Appendix~\ref{EulerSingle}, the Euler transformation leaves it unchanged. For
$\e=0.3$ the loss of the localized character of the $S_5$ solution is 
considerably more pronounced, and it begins to
appear in the Euler approximations as well, since the approximation of the outermost site is common
to all of them. Finally, for $\e=0.5$ the loss of localization propagates to the inner sites of the
$E_5$ and $E_{5i}$ approximations. Note that, to the naked eye, $E_5$ and $E_{5i}$ are
indistinguishable, so that a more quantitative diagnostic
is needed to discern between the two.

\begin{figure}
\includegraphics[width=0.5\textwidth]{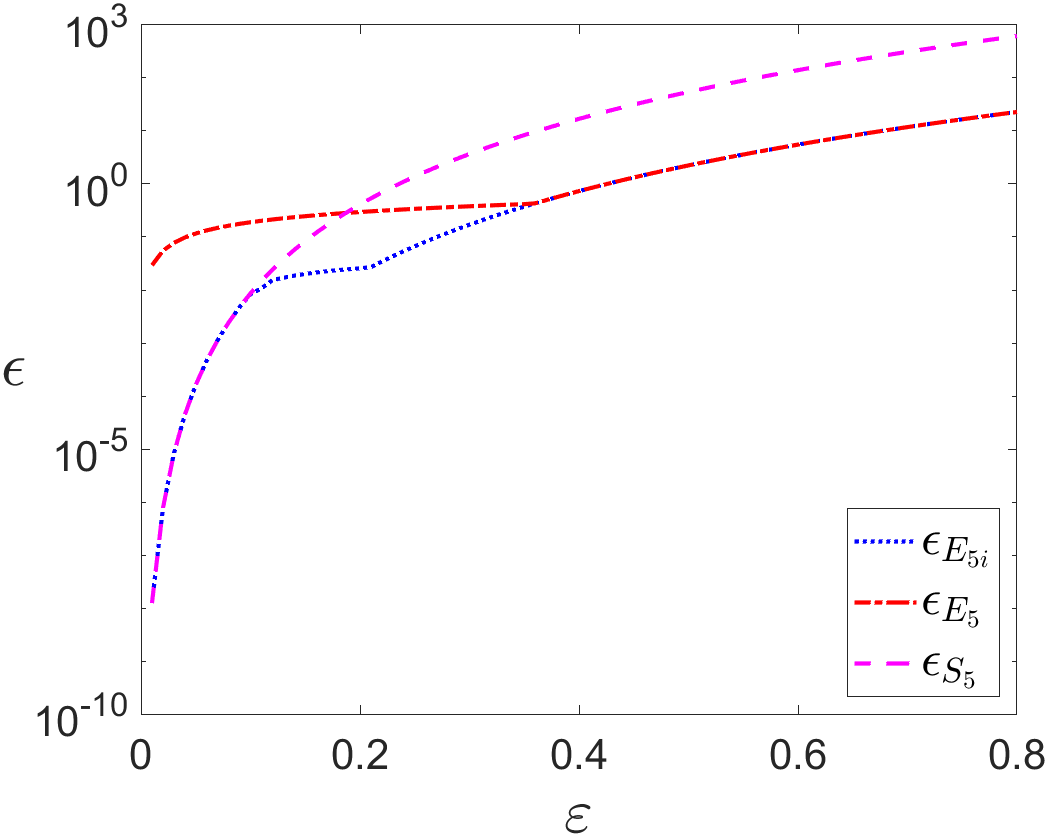}
\caption{Absolute 
relative errors, $\epsilon_{E_{5i}}$ (blue, dotted) of the improved Euler $E_{5i}$, $\epsilon_{E_5}$ (red,dash-dotted) of the Euler $E_5$ and $\epsilon_{S_5}$ (magenta, dashed) of the standard series $S_5$ as percentages of the approximate solutions on a semilogarithmic scale, for the
single-site configuration. The improved Euler approximation consistently exhibits the smallest error with respect to the numerical solution over the range of $\e$ considered.}
\label{on_site_error}
\end{figure}

In order to quantify the deviation of the approximate solutions from the numerical one, we introduce
the relative $\ell_{\infty}$ error
\begin{equation}
\relerr_{S_i}=100~ \frac{\|S_i-N\|_{\infty}}{\|N\|_{\infty}},
\label{error}
\end{equation}
expressed as a percentage; the errors $\relerr_{E_5}$ and $\relerr_{E_{5i}}$ associated with the
$E_5$ and $E_{5i}$ approximations are defined analogously. Figure~\ref{on_site_error} compares
$\relerr_{S_5}$, $\relerr_{E_5}$ and $\relerr_{E_{5i}}$. The improved Euler approximation yields the
smallest error among all the approximations considered: for small $\e$ it reproduces the error of
the fifth-order series $S_5$, whereas for larger $\e$ it follows that of $E_5$. In other words, the
selection rule of Table~\ref{tab:Eulers} automatically switches between the two limiting behaviours,
so that the improved Euler transformation provides the closest agreement with the numerical solution
over the entire range of $\e$ considered. This mechanism will, in fact,
be found to recur in the
two remaining configurations.

This behavior reveals that, for small enough values of $\varepsilon$, the most accurate series is the standard $S_5$ one, while the power of the Euler transformation is revealed for larger values of $\varepsilon$. Thus, as a general remark we could say that, if someone wants the most refined approximation of the solution, he should use the improved Euler transformation. But this comes with the cost of a more complex algorithm. In order to avoid these complications, he can use the standard series for values up to $\varepsilon\leq0.2$ and  switch to the standard Euler transformation for  higer values of $\varepsilon$. This improves significantly the convergence of the solution providing an error of $\approx1\%$ for values of $\e$ up to $0.5$ which is adequate for most uses.


\subsection{Two-site configuration: the in-phase case}

To obtain the series approximation of the in-phase two-site configuration we again consider an
expansion of degree $n$ of the form~\eqref{exp0}, now for the configuration
$$S_n=\{\ldots, \vi_{2}, \vi_{1}, \vi_{0}, \vi_{0}, \vi_{1}, \vi_{2},\ldots\},$$
so that in the anti-continuum limit we recover the trivially localized configuration
$\{\ldots,0,0,1,1,0,0,\ldots\}$. Inserting this approximation into~\eqref{algebraic} and applying
order-by-order perturbation theory, we obtain, for $n=5$,
\beq\begin{array}{rclcrcl}
\vi_{0}& \simeq &1+\dfrac{ \e}{2}-\dfrac{5 \e^{2}}{8}+\dfrac{21 \e^{3}}{16}-\dfrac{493 \e^{4}}{128}+\dfrac{3391 \e^{5}}{256}  &\qquad\qquad &\vi_{3}&\simeq&\e^{3}-\dfrac{11 \e^{4}}{2}+\dfrac{195 \e^{5}}{8},\\[0.2cm]
\vi_{1}& \simeq &\e-\dfrac{3 \e^{2}}{2}+\dfrac{35 \e^{3}}{8}-\dfrac{247 \e^{4}}{16}+\dfrac{7587 \e^{5}}{128} & &
\vi_{4}& \simeq &\e^{4}-\dfrac{15 \e^{5}}{2},\\[0.2cm]
\vi_{2}& \simeq &\e^{2}-\dfrac{7 \e^{3}}{2}+\dfrac{99 \e^{4}}{8}-\dfrac{731 \e^{5}}{16}& &
\vi_{5}& \simeq &\e^{5}.
\label{in_phase_dnls}
\end{array}\eeq
Applying the standard and the improved Euler transformations to $S_5$ yields the expansions which are listed in the document that supplements this manuscript.

\begin{figure}[!b]
\includegraphics[width=0.32\textwidth]{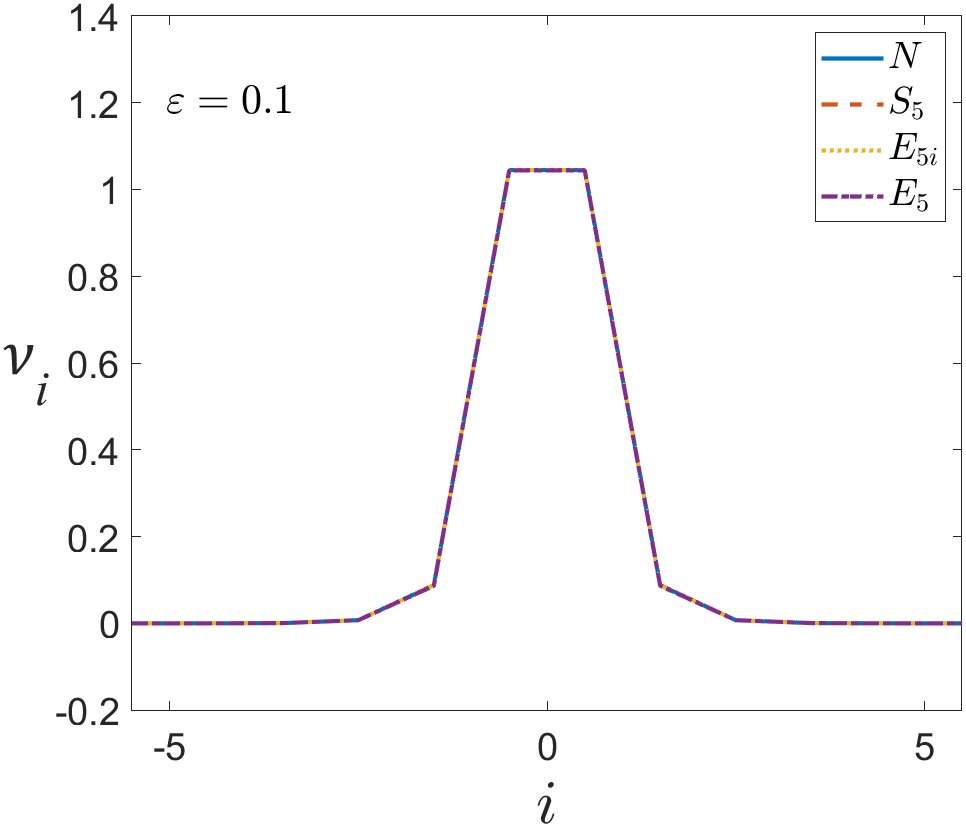}
\includegraphics[width=0.32\textwidth]{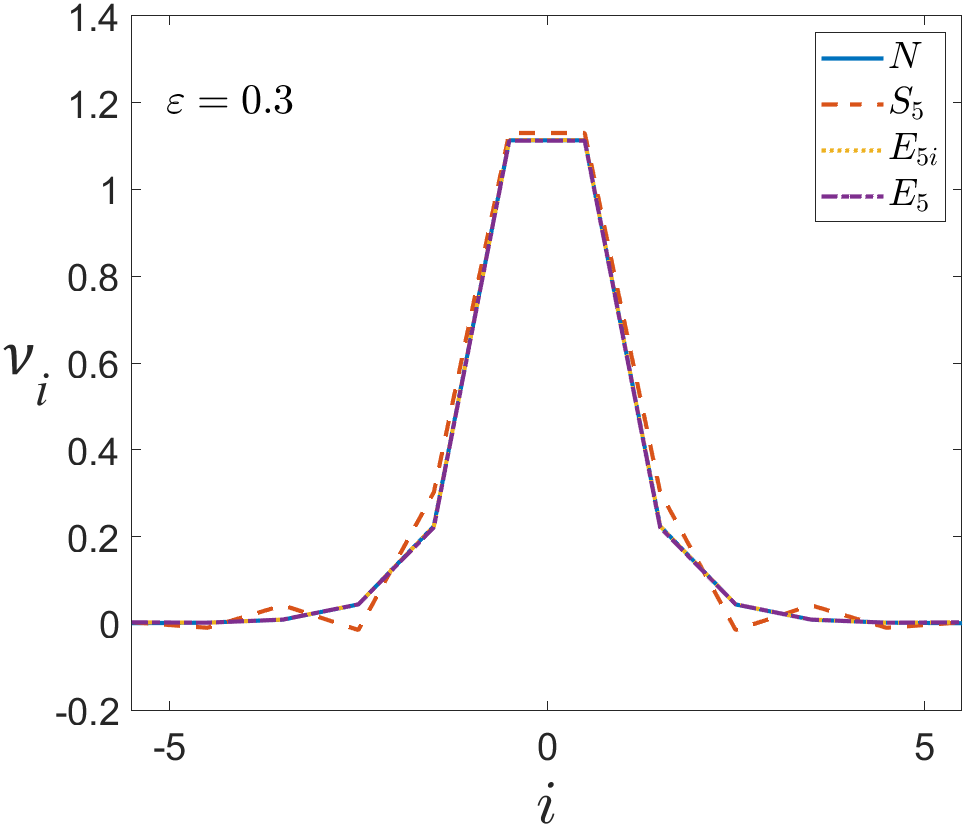}
\includegraphics[width=0.32\textwidth]{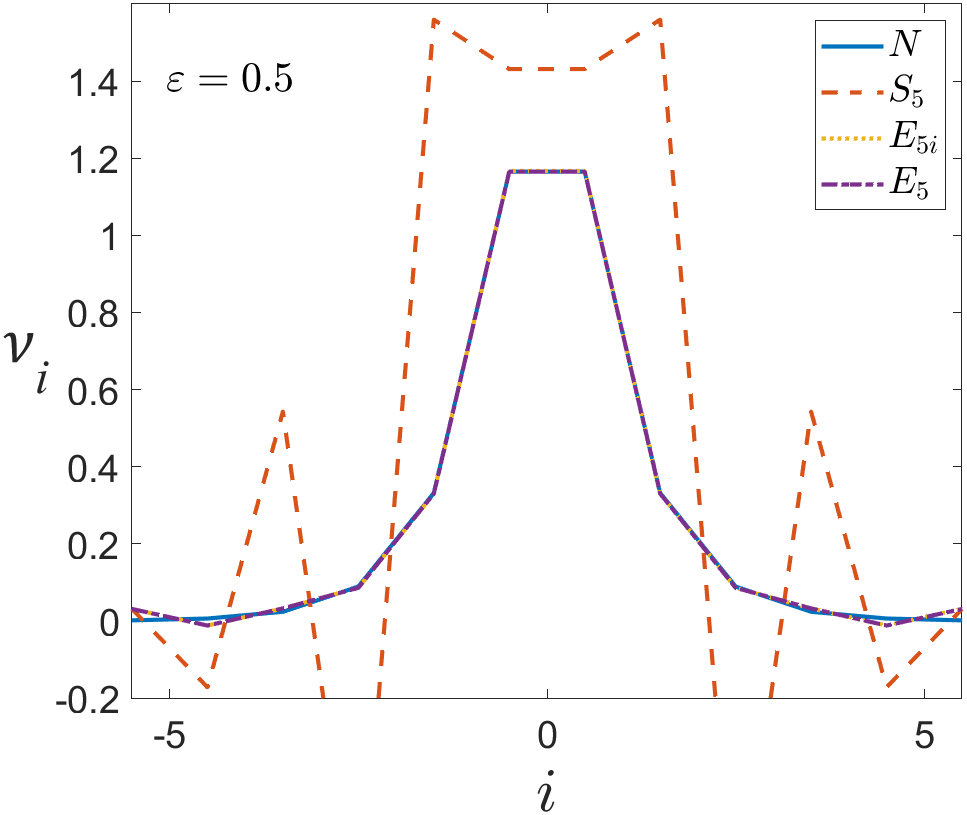}
\caption{Profile of the in-phase discrete-soliton solution on a linear scale, for three values of
the coupling parameter, namely $\e=0.1, 0.3, 0.5$. The numerical solution N (blue, solid) is compared with the standard $n=5$
term series expansion $S_5$ (red, dashed), the Euler transform result ${E}_5$ (purple, dash-dotted) and
the improved variant the latter ${E}_{5i}$ (yellow, dotted).}
\label{fig:portrait_in_phase}
\end{figure}

Figure~\ref{fig:portrait_in_phase} shows the profile of the numerically computed discrete soliton $N$
together with the $S_5$, $E_5$ and $E_{5i}$ approximations for the in-phase case. In this configuration we have two central oscillators and the left part of the lattice moves in-phase with its right one, i.e., the
configuration preserves the parity symmetry throughout the
relevant parametric continuation (as it is a property of
the original DNLS model). The situation described in Figure~\ref{fig:portrait_in_phase} is quite similar to what we observed in the single-site configuration. For low coupling ($\e=0.1$) all the
series provide a good approximation. For larger values of $\e$, the Euler approximations still provide a good approximation while the original series becomes quantitatively irrelevant.

\begin{figure}
\includegraphics[width=0.32\textwidth]{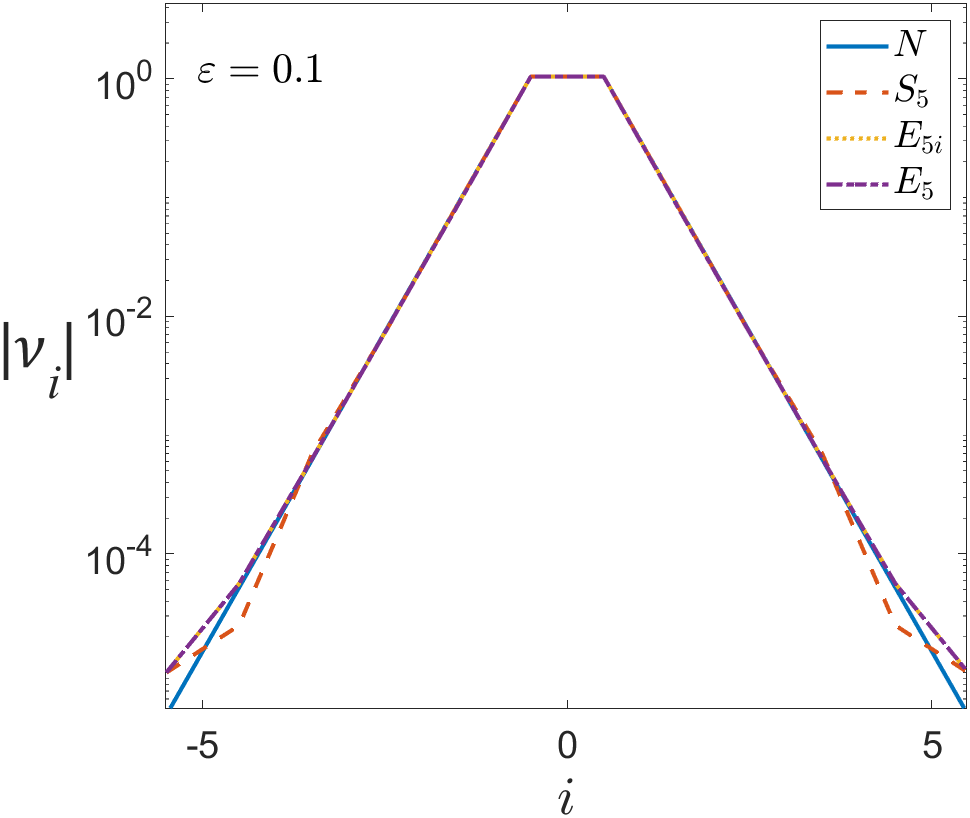}
\includegraphics[width=0.32\textwidth]{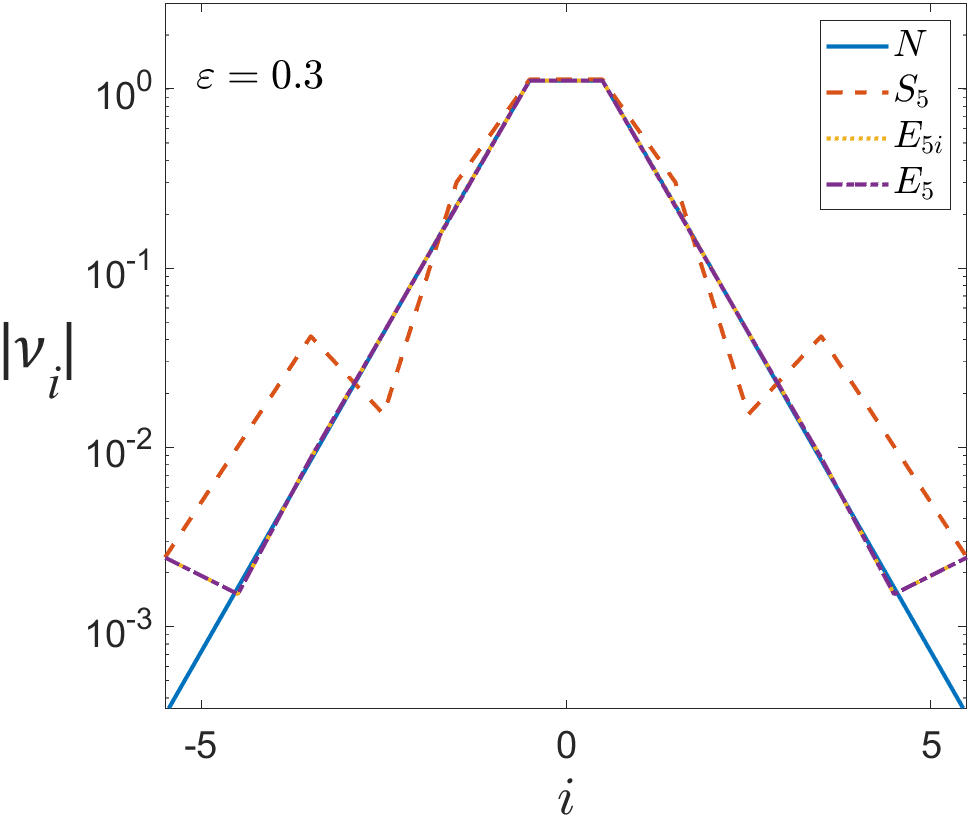}
\includegraphics[width=0.32\textwidth]{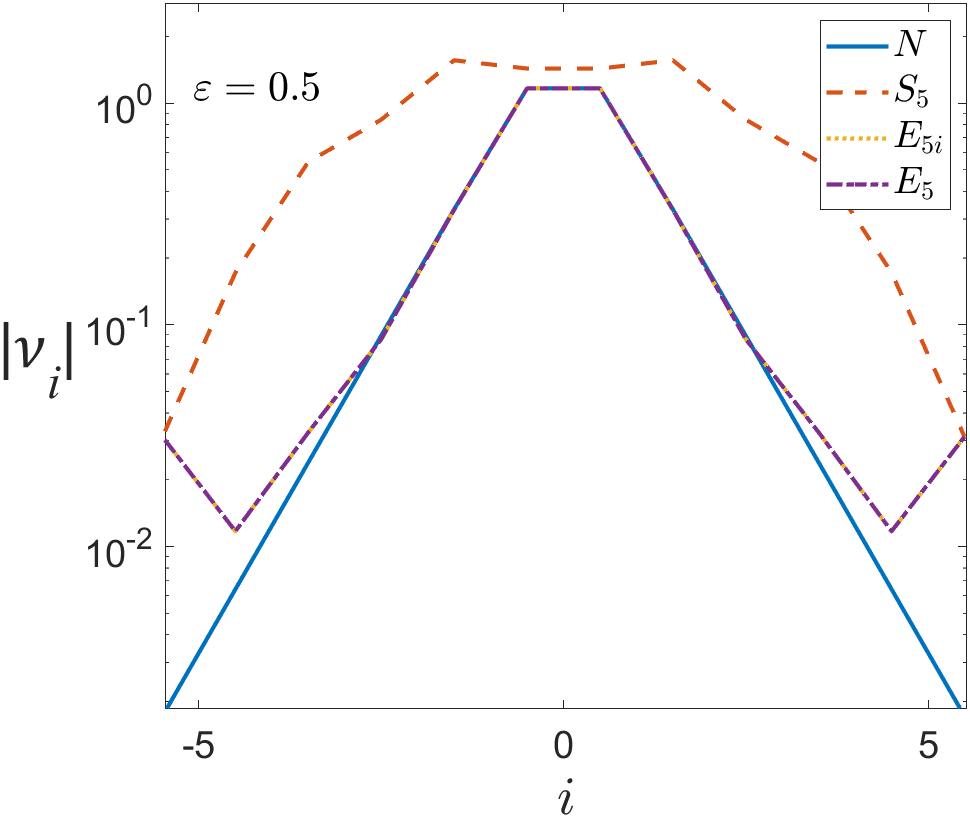}
\caption{Profile of the in-phase discrete-soliton solution on a semilogarithmic scale, for three
values of the coupling parameter, namely $\e=0.1, 0.3, 0.5$. The numerical solution N (blue, solid) is compared with the standard $n=5$
term series expansion $S_5$ (red, dashed), the Euler transform result ${E}_5$ (purple, dash-dotted) and
the improved variant the latter ${E}_{5i}$ (yellow, dotted).} 
\label{fig:portrait_in_phase_log}
\end{figure}

In the semilogarithmic version of the profiles
(Fig.~\ref{fig:portrait_in_phase_log}), the deviations at the edges of the lattice are already discernible at small values of $\e$, while for larger values of $\e$ we can observe the loss of the localization of the $S_5$ solution. On the other hand, for large values of $\e$, while it is obvious that the $S_5$ solution 
is now highly inaccurate, we can also observe that also the delocalization of the $E_5$ and $E_{5i}$ solutions becomes gradually evident.

\begin{figure}
\includegraphics[width=0.5\textwidth]{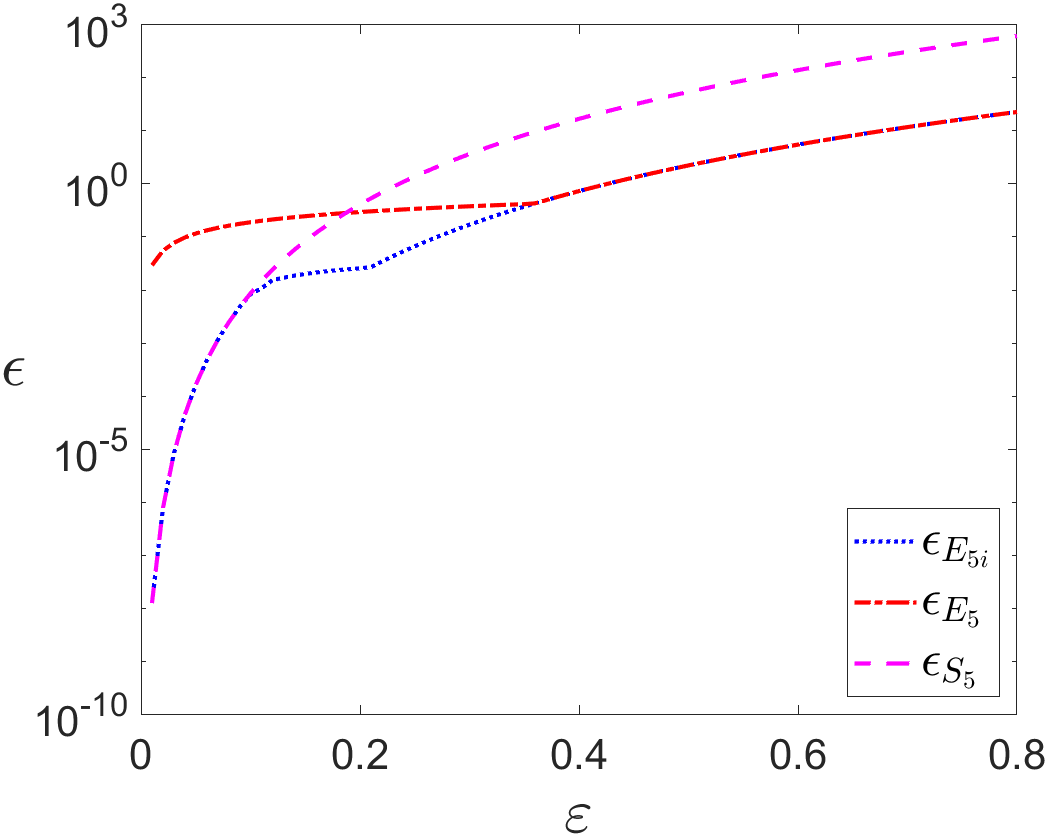}
\caption{Absolute relative errors as percentages of the approximate solutions on a semilogarithmic scale, for the in-phase
configuration. Here $\relerr_{S_5}$ (magenta, dashed), $\relerr_{E_5}$ (red, dash-dotted) and $\relerr_{E_{5i}}$ (blue, dotted) denote the errors of the
standard series $S_5$, the Euler $E_5$ and the improved Euler $E_{5i}$ approximations, respectively,
with respect to the numerical solution.}
\label{in_phase_error}
\end{figure}

The accuracy of the resulting approximations, measured by the percentage of the absolute
relative error of Eq.~\eqref{error}, is shown in Fig.~\ref{in_phase_error}. The behavior is the same as in
the single-site case: the improved Euler approximation tracks $S_5$ at small $\e$ and $E_5$ at large
$\e$, and therefore remains the most accurate of the three throughout the
relevant continuation.

\subsection{Two-site configuration: the out-of-phase case}
For the out-of-phase two-site configuration we consider the same expansion~\eqref{exp0} applied to
$$S_n=\{\ldots, +\vi_{2}, +\vi_{1}, +\vi_{0}, -\vi_{0}, -\vi_{1}, -\vi_{2},\ldots\},$$
so that the anti-continuum limit is the trivially localized configuration
$\{\ldots,0,0,1,-1,0,0,\ldots\}$. Proceeding as before for $n=5$, we obtain
\beq\begin{array}{rclcrcl}
\vi_{0}& \simeq &1+\dfrac{ 3\e}{2}-\dfrac{13 \e^{2}}{8}+\dfrac{55 \e^{3}}{16}-\dfrac{1213 \e^{4}}{128}+\dfrac{7501 \e^{5}}{256}  &$ $\qquad$ $ &\vi_{3}& \simeq &\e^{3}-\dfrac{9 \e^{4}}{2}+\dfrac{139 \e^{5}}{8},\\[0.2cm]
\vi_{1}& \simeq &\e-\dfrac{ \e^{2}}{2}+\dfrac{11 \e^{3}}{8}-\dfrac{53 \e^{4}}{16}+\dfrac{1203 \e^{5}}{128} & &
\vi_{4}& \simeq &\e^{4}-\dfrac{13 \e^{5}}{2},\\[0.2cm]
\vi_{2}& \simeq &\e^{2}-\dfrac{5 \e^{3}}{2}+\dfrac{59 \e^{4}}{8}-\dfrac{361 \e^{5}}{16}& &
\vi_{5}& \simeq &\e^{5}.
\label{out_of_phase_dnls}
\end{array}\eeq

\begin{figure}
\includegraphics[width=0.5\textwidth]{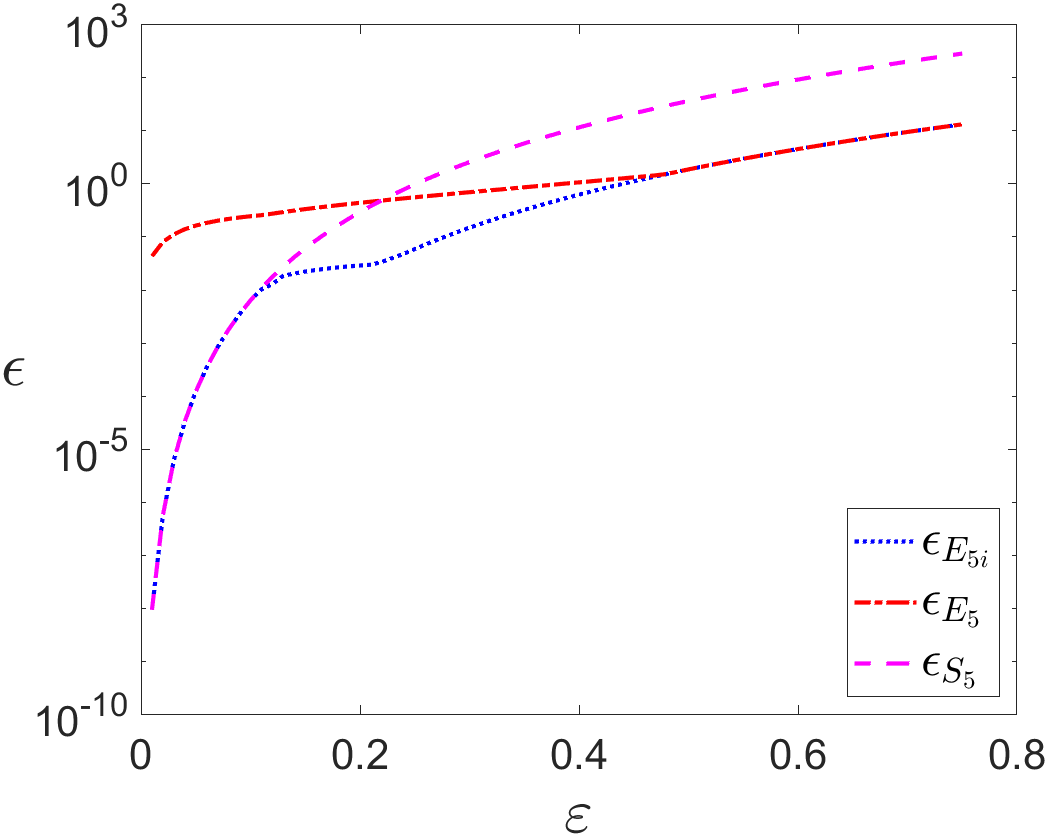}
\caption{Absolute relative errors as percentages of the approximate solutions on a semilogarithmic scale, for the
out-of-phase configuration. Here $\relerr_{S_5}$ (magenta,dashed), $\relerr_{E_5}$ (red,dash-dotted) and $\relerr_{E_{5i}}$ (blue,dotted) denote the
errors of the standard series $S_5$, the Euler $E_5$ and the improved Euler $E_{5i}$ approximations,
respectively, with respect to the numerical solution.}
\label{out_of_phase_error}
\end{figure}

The corresponding transformed expansions are provided in the supplementary material of this manuscript. This configuration is a two-site state
(at the AC limit) where half of the right part of the lattice moves in opposition with the left one. In this case we chose not to present portrait figures to avoid repetition since the findings are quite similar to the ones shown in Figs.\ref{fig:portrait_in_phase}--\ref{fig:portrait_in_phase_log}.  The
associated percentage absolute relative errors are shown in Fig.~\ref{out_of_phase_error}. Once more the improved Euler
approximation interpolates between $S_5$ at small $\e$ and $E_5$ at large $\e$, and is the most
accurate of the three over the whole range of $\e$ considered. Note that the calculations here have been done up to $\e\simeq0.77$ due to the specific range of existence of the solution in the out-of-phase configuration. At this specific coupling value, the linear spectrum reaches 0 and the localized solution is terminated by bifurcating to a non-localized one. 

It is worth explaining why convergence acceleration is needed in the first place.
We have computed the expansion of $\nu_0$ to order $34$ (results
not shown) and examined the ratios
of successive coefficients. For all three configurations, these ratios remain bounded,
with
\[
3.6 \lesssim
\left|
\frac{\nu_0^{(k)}}{\nu_0^{(k-1)}}
\right|
\lesssim 5.1,
\]
rather than exhibiting the approximately linear growth with $k$ characteristic of
factorially divergent asymptotic expansions. Although finite-order calculations cannot
establish convergence, the observed behavior strongly suggests that the expansions have
a finite, nonzero estimated radius of convergence of order $\epsilon_c \simeq 0.2$.
As $\epsilon$ approaches this value, successive terms cease to decrease rapidly, so a
truncation after only five terms becomes increasingly inaccurate. This is consistent with
the deterioration of the $S_5$ approximation observed in
Figs.~\ref{on_site_error}--\ref{out_of_phase_error}, and motivates the use of the Euler
transformation.

\section{Stability of the two-site configurations}\label{sec:stability}
As a measure of the agreement of our approximations we now examine the linear stability of the computed
solutions, which is determined by the eigenvalues of the stability matrix of the cubic DNLS equation
(see e.g.\ Ref.~\cite{Kevrekidis2009}\S 2.1),
\beq
J=
\left(
\begin{array}{cc}
0 & A_{1} \\
A_{2} & 0\\ \end{array} \right),
\label{eq:stability_matrix}\eeq
where, for a lattice of sites $j,k=-M,\ldots,M$, the two blocks are the tridiagonal matrices
\begin{equation}
(A_{1})_{jk}=\left( 1+2\e- \vi_{j} ^{2}\right)\delta_{jk}-\e\left( \delta_{j,k+1}+\delta_{j,k-1}\right),\qquad
(A_{2})_{jk}=-\left( 1+2\e-3 \vi_{j} ^{2}\right)\delta_{jk}+\e\left( \delta_{j,k+1}+\delta_{j,k-1}\right),
\label{eq:blocks}
\end{equation}
for the real waveforms considered herein, with $\delta_{jk}$ the Kronecker delta.

As is discussed, e.g., in~\cite{Kevrekidis2009}, the single site configuration
is not relevant to consider in the vicinity of the AC limit (but rather
only in the vicinity of the continuum one), as there are no nonzero
eigenvalues that are bifurcating from the origin of the
spectral plane, i.e., $\lambda=0$. Hence, we focus on the
in-phase and out-of-phase two-site configurations. 

\subsection{The in-phase configuration}
As shown in Ref.~\cite{PelinovskyKevrekidisFrantzeskakis2005a}, in the in-phase two-site case the
linear spectrum of the discrete-soliton solution lies on the imaginary axis in two conjugate bands,
while a pair of eigenvalues remains at zero, as is standard for a 
standing wave in the DNLS model (which enjoys the so-called phase or gauge
invariance associated with the conservation of the $l^2$ norm). In addition, an
isolated pair of opposite eigenvalues $(\lambda,-\lambda)$ appears on the real axis, the positive
member of which renders the solution unstable. To leading order this unstable eigenvalue was found
in Ref.~\cite{PelinovskyKevrekidisFrantzeskakis2005a} to be $\lam{1/2}=2\sqrt{\e}$, where the
superscript records the highest power of $\e$ retained.

\begin{figure}
\includegraphics[width=0.45\textwidth]{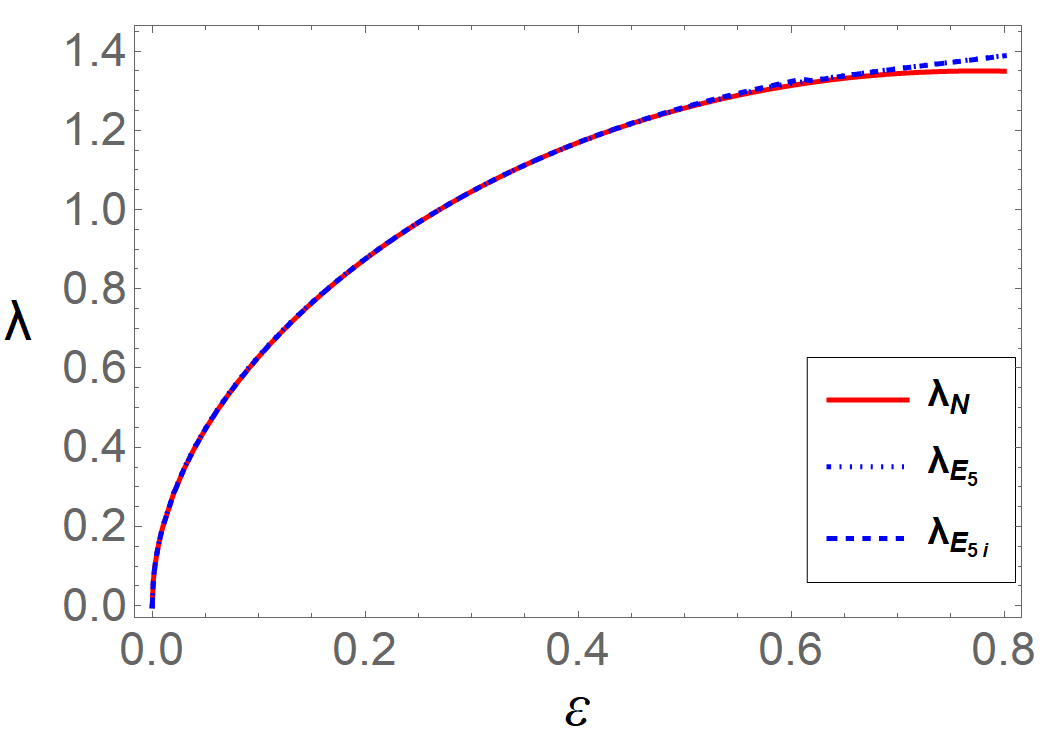}
\includegraphics[width=0.45\textwidth]{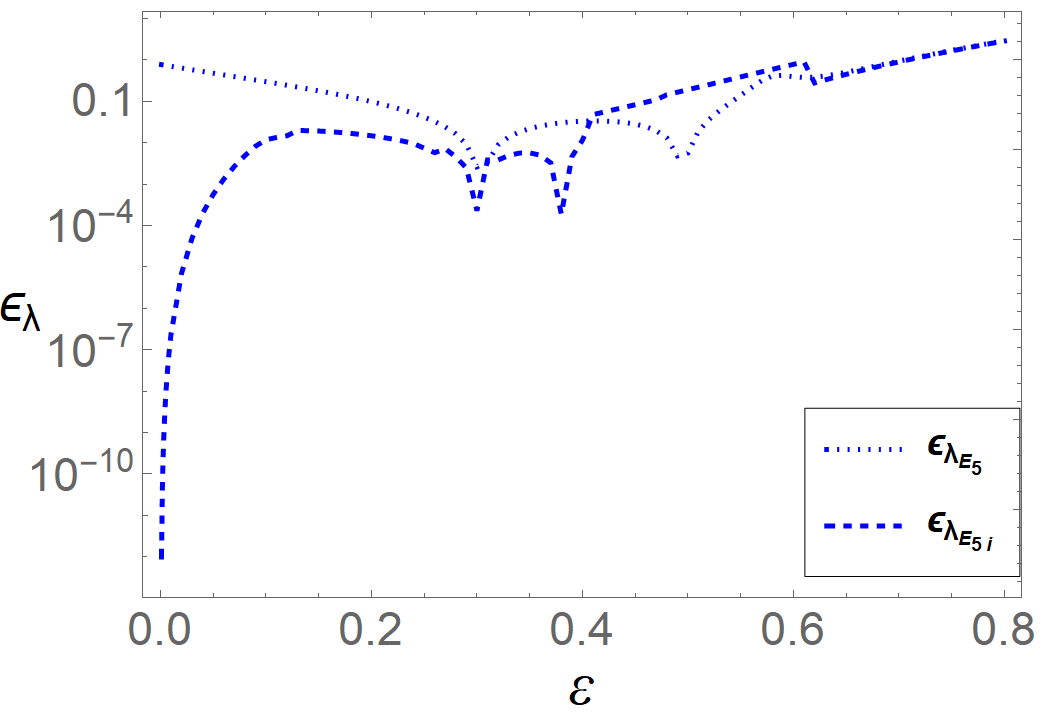}
\caption{Comparison between the standard $\lambda_{E_5}$ (Blue, dotted)- and the improved-Euler $\lambda_{E_{5i}}$ (Blue, dashed), obtained by the semi-numerical method described in the text, with the numerical $\lambda_N$ (red, solid) unstable (real) eigenvalue for the in-phase
case. The left panel shows an excellent  agreement between all curves, which is confirmed by the
relative errors $\relerr_{\lambda_{E_{5}}}$ and $\relerr_{\lambda_{E_{5i}}}$, plotted on a semilogarithmic scale in the right panel.}
\label{euler_improved}
\end{figure}

We wish to calculate the unstable eigenvalue by using a semi-numerical approach  and the improved Euler
$E_{5i}$ approximation of this configuration. The full transformation is presented in the supplementary material of this manuscript. 
To this end, the numerical values of the transformed solutions are substituted into the stability matrix for each
$\varepsilon\in[0,0.8]$, and the corresponding eigenvalue, denoted by $\lambda_{E_{5i}}$, is computed. Then we repeat the procedure in order to calculate the same eigenvalue but by using the $E_5$ approximation. We call this eigenvalue $\lambda_{E_5}$. The results are shown in the left panel of Figure~\ref{euler_improved}, where these eigenvalues are depicted together with $\lambda_N$ which is the eigenvalue obtained from the fully numerical solution. The curves are almost indistinguishable up to $\e=0.6$ while they are still really close up to $\e=0.8$. To examine the difference more closely we compute the absolute relative error of $\lambda_{E_{5}}$ and $\lambda_{E_{5i}}$, taking
$\lambda_N$ as the reference value. The error for $\lambda_{E_{5}}$ is defined as
\begin{equation}
 \relerr_{\lambda_{E_{5}}}=100~ \frac{\left| \lambda_{E_{5}}-\lambda_{N}\right|}{\left| \lambda_N \right|},
 \label{a.r.e}
\end{equation}
while $\epsilon_{\lambda_{E_{5i}}}$ is defined similarly. The result is shown in the right panel of Fig.~\ref{euler_improved}. As we can see the error remains below $1\%$ even for $\e=0.7$. We can see again that the $E_{5i}$ solution provides better results for small $\e$ but in general, as we mentioned also in the solution section, the standard Euler transform is adequate for most uses while the improved one is useful if fine tuning is needed.

Although the procedure above delivers the unstable eigenvalue, it does not provide a closed-form
expression for it as a function of the coupling parameter, of the kind obtained in
Ref.~\cite{PelinovskyKevrekidisFrantzeskakis2005a}. We therefore turn to an asymptotic expansion of
the unstable eigenvalue in powers of $\e$, with the aim of deriving a more tractable analytical
approximation.

Following a procedure analogous to the one described in the previous section, we compute the
approximate solutions $S_n$ for $n=1,\ldots,5$, insert them into the stability matrix and evaluate
the corresponding unstable eigenvalue analytically up to order $\e^{9/2}$ using \textsc{Mathematica}.
The results are collected in Table~\ref{tab:eig_inphase}.

\begin{table}[htb]
\renewcommand{\arraystretch}{2.3}
$$\begin{array}{|c|c|}
\hline
S_n & \text{Unstable eigenvalue}\\
\hline
S_1 & \sqrt{\e }\left( 2-\dfrac{5}{8}\e +\dfrac{455}{256}\e ^{2}-\dfrac{38877}{4096}\e ^{3}+\dfrac{19659275}{262144}\e^{4} \right)\\[0.5cm]
\hline
S_2 & \sqrt{\e }\left( 2+\dfrac{5}{16}\e ^{2}-\dfrac{1425}{128}\e ^{3}+\dfrac{80639}{1024}\e^{4}\right)\\[0.5cm]
\hline
S_3 & \sqrt{\e }\left( 2-\e ^{2}-\dfrac{1005}{128}\e ^{3}+\dfrac{661}{8}\e^{4}\right)\\[0.5cm]
\hline
S_4 & \sqrt{\e }\left( 2-\e ^{2}-4\e ^{3}+\dfrac{18687}{256}\e^{4}\right)\\[0.5cm]
\hline
S_5 & \sqrt{\e }\left( 2-\e ^{2}-4\e ^{3}+\dfrac{239}{4}\e^{4}\right)\\[0.5cm]
\hline
\end{array}$$
    \caption{The unstable eigenvalue of the in-phase configuration, obtained from the solution
    approximations $S_1$ through $S_5$.}
    \label{tab:eig_inphase}
\end{table}

\begin{figure}
\includegraphics[width=0.45\textwidth]{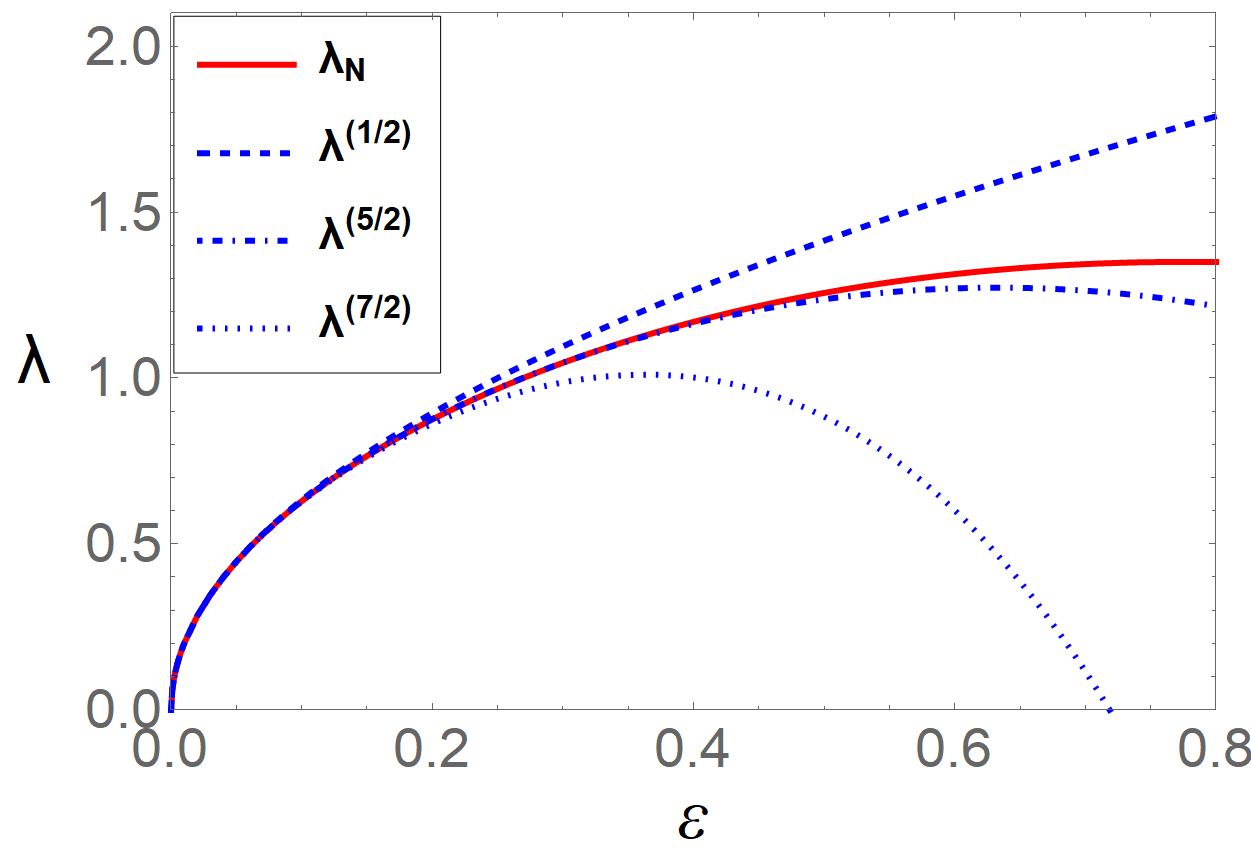}
\caption{Comparison of the asymptotic approximations of the unstable eigenvalue in the in-phase
case, $\lam{1/2}$ (blue, dashed), $\lam{5/2}$ (blue, dash-dotted) and $\lam{7/2}$ (blue, dotted), with the numerical eigenvalue $\lambda_N$ (red,solid). The
$\lam{5/2}$ approximation provides the best agreement with the numerical result.}
\label{in_phase_li}
\end{figure}

Table~\ref{tab:eig_inphase} shows clearly how the coefficients of the eigenvalue expansion 
stabilize
as the order of the solution increases. The leading-order result
$\lam{1/2}=2\e^{1/2}$ of Ref.~\cite{PelinovskyKevrekidisFrantzeskakis2005a} is already recovered
from the first-order approximation of the solution. The coefficient of the $\e^{5/2}$ term
stabilizes at the level of the $S_3$ approximation, giving
$\lam{5/2} = 2\e^{1/2} - \e^{5/2}$, while the coefficient of the $\e^{7/2}$ term requires the $S_4$
solution and yields $\lam{7/2} = 2\e^{1/2} - \e^{5/2} - 4\e^{7/2}$. These approximations are
compared with the numerical eigenvalue $\lambda_N$ in Fig.~\ref{in_phase_li}. The $\lam{5/2}$
approximation is clearly the closest to the numerical result; including the next term, of order
$\e^{7/2}$, makes the approximation worse and causes it to deviate from $\lambda_N$. This behaviour
is characteristic of an asymptotic rather than a convergent expansion: the optimal truncation is
attained at order $\e^{5/2}$, beyond which further terms degrade the approximation.

\begin{figure}
\includegraphics[width=0.45\textwidth]{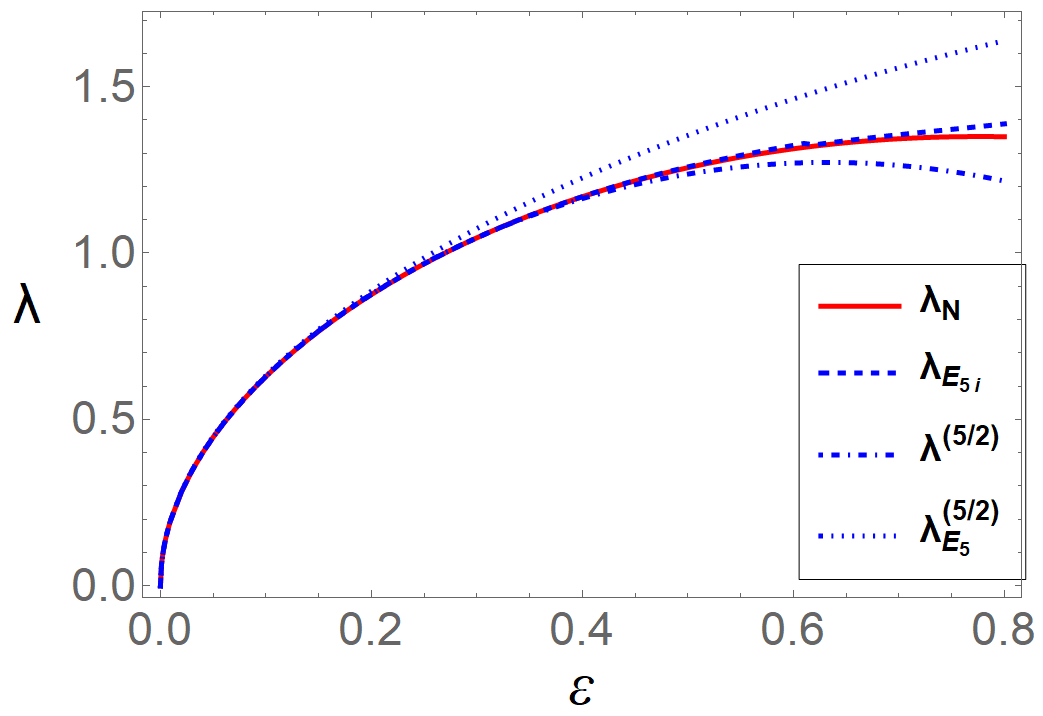}
\includegraphics[width=0.45\textwidth]{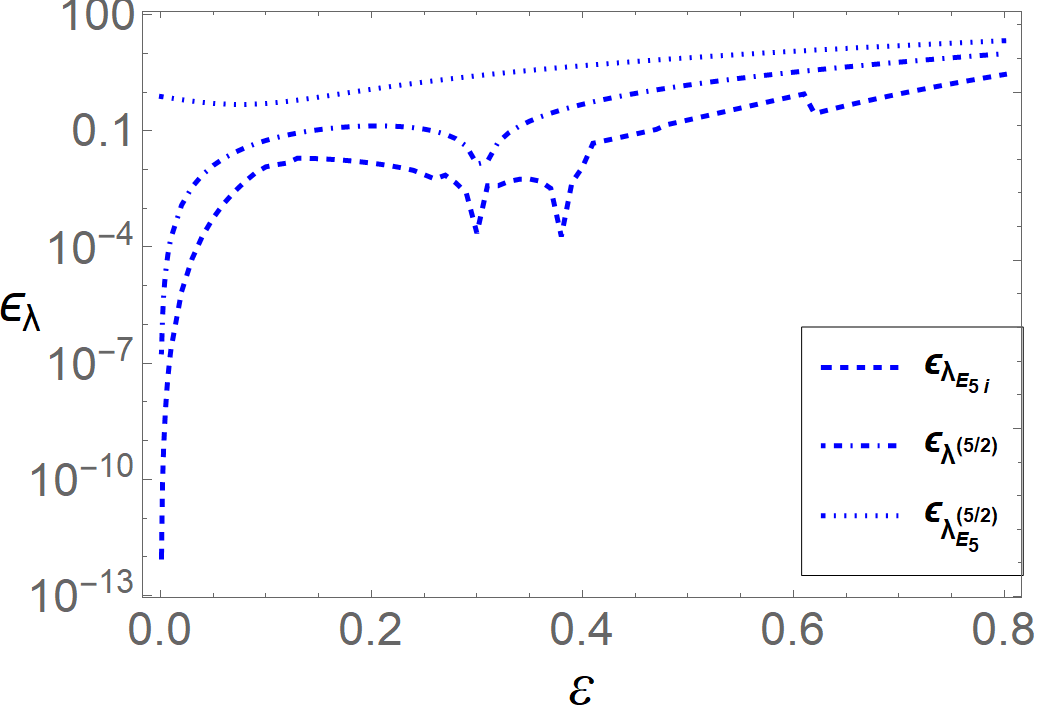}
\caption{
The unstable eigenvalue of the in-phase configuration. The left panel compares the
numerical eigenvalue $\lambda_N$ (red,solid), the optimal truncation $\lam{5/2}$ (blue, dash-dotted), the result $\lambda^{(5/2)}_{E_5}$ (blue, dotted)
obtained with the standard Euler transform and truncated to terms up to $\e^{5/2}$, and the result $\lambda_{E_{5i}}$ (blue, dashed) obtained from the
semi-numerical use of the improved Euler transform. The right panel shows the relative error of each
approximation with respect to the numerical eigenvalue.}
\label{in_phase_sigrisi}
\end{figure}

Finally, we apply the standard Euler transform $E_5$ to the $S_5$ solution, insert the transformed
profile into the stability matrix and obtain, up to ${\cal O}(\e^{5/2})$, the approximation
$$\lambda^{(5/2)}_{E_5}= \dfrac{\sqrt{\e }}{\sqrt{65}}\left( \dfrac{65}{4}-\dfrac{1287}{1024} \e - \dfrac{1955013}{2621440} \e ^{2}\right).$$
This approximation is shown together with $\lambda_{E_{5i}}$, $\lam{5/2}$ and $\lambda_N$ in the
left panel of Fig.~\ref{in_phase_sigrisi}, while the right panel displays the corresponding relative
errors with respect to the numerical eigenvalue on a semilogarithmic scale.

The improved Euler transform provides the best agreement with the numerical solution, at the cost of
a comparatively cumbersome implementation and without providing an analytic formula. The $\lam{5/2}$ approximation, although less accurate,
offers a considerably simpler closed-form expression and may therefore be preferable whenever high
accuracy is not required. It is an intriguing observation
(for which we have no intuitive explanation) that the error
of $\lam{5/2}$ is smaller than $\lambda^{(5/2)}_{E_5}$.

\subsection{The out-of-phase configuration}
In this case, it was already established in~\cite{PelinovskyKevrekidisFrantzeskakis2005a} that
an isolated pair of eigenvalues also exists yet in this case it lies on the imaginary axis for sufficiently small
$\e$. Accordingly, the configuration is spectrally stable until the pair collides with the linear spectrum
and the solution is destabilized through a Hamiltonian--Hopf bifurcation. 
However, 
it is still instructive to track
the relevant eigenvalues, both as a further test of our approach and in order to see whether the estimate of
Ref.~\cite{PelinovskyKevrekidisFrantzeskakis2005a} can be improved. We therefore repeat the
procedure of the previous subsection: The full improved Euler transformation $E_{5i}$ is calculated 
(and presented as supplementary material of the manuscript). Then, the corresponding transformed solutions are evaluated
numerically for each $\e\in[0,0.6]$, inserted into the stability
matrix~\eqref{eq:stability_matrix}--\eqref{eq:blocks}, and the isolated eigenvalue with positive
imaginary part is computed. We do not pursue larger values of $\e$, 
since there are several bifurcations occurring rendering the results 
irrelevant as we are going to see in what follows.

\begin{figure}
\includegraphics[width=0.45\textwidth]{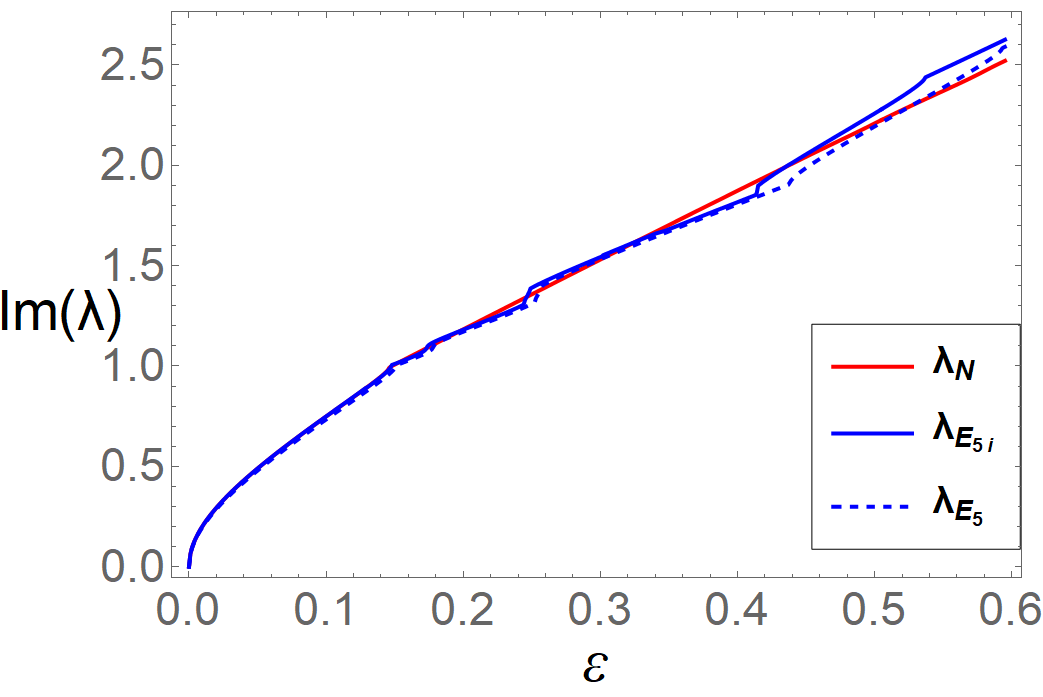}\hspace{0.2cm}
\includegraphics[width=0.45\textwidth]{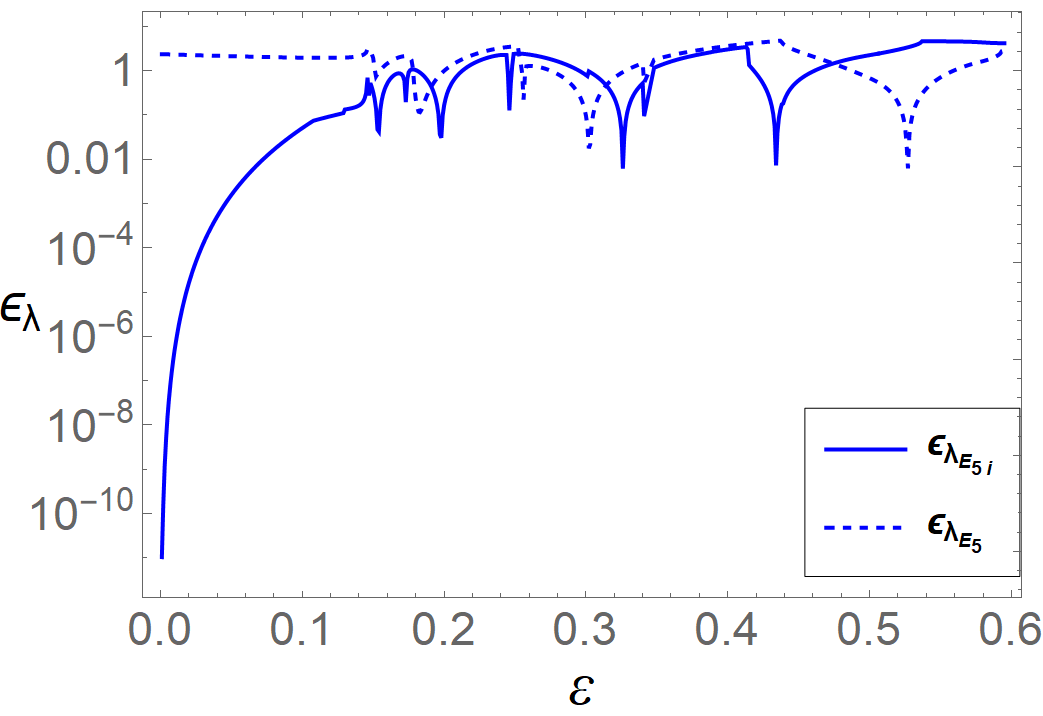}
\caption{The out-of-phase configuration. The left panel shows $\lambda_{E_{5}}$ (blue, dashed) and $\lambda_{E_{5i}}$ (blue,solid), obtained by the semi-numerical method described in the text, together with
$\lambda_{N}$ (red,solid), while the right panel shows the corresponding relative error
$\relerr_{\lambda_{E_{5}}}$ and $\relerr_{\lambda_{E_{5i}}}$.}
\label{out_euler_improved}
\end{figure}

The results are shown in the left panel of Fig.~\ref{out_euler_improved}, where the imaginary part
of the $\lambda_{E_{5}}$ and $\lambda_{E_{5i}}$ eigenvalues is compared with its fully numerical counterpart
$\lambda_N$. The agreement between the curves is adequate, but neither as close nor as smooth as
in the in-phase configuration (cf.\ Fig.~\ref{euler_improved}). The reason is that here the isolated
eigenvalue collides with the linear spectrum, giving rise to a complex quartet~\cite{PelinovskyKevrekidisFrantzeskakis2005a}, and consequently
the  ``interaction'' between the relevant eigenvalues affects
such collisions and the resulting eigenvalue dependencies. For the fully numerical
eigenvalue this happens only once, at $\e\simeq0.146$, producing a single spike in the
$\lambda(\e)$ curve: on a sufficiently large lattice the complex pair cannot re-enter the imaginary
axis before it reaches the end of the linear spectrum. The series approximations, by contrast, have
only $12$ non-zero sites, so that the corresponding stability matrix is $24\times24$ and possesses
only $10$ eigenvalues with positive imaginary part. There is consequently ample room on the
imaginary axis for the complex pair to leave and re-enter repeatedly, which produces the succession
of ``corners'' visible in $\lambda_{E_{5i}}(\e)$. We emphasize that this is an artifact of the
finite truncation of the series solution, not of the acceleration procedure itself. The right panel
of the same figure shows the errors $\relerr_{\lambda_{E_{5}}}$ and $\relerr_{\lambda_{E_{5i}}}$ defined in Eq.~\eqref{a.r.e}. Several
spikes are visible, caused both by the collisions described above and by the switching between the
branches of the improved Euler transformation, but the error nevertheless remains of the order of
$1\%$.
At this point one could also be tempted to compare the real part of the eigenvalues as well.
However, this is not very meaningful due to the many departures and re-entrances of the eigenvalue  from the imaginary axis as described above. 
On the other hand, it is interesting to observe that this semi-numerical approach is very accurate. 
Remarkably, it reveals rather accurately the bifurcation point
in which the linearly stable solution becomes unstable.

As before, since this semi-numerical approach  does not yield a single closed-form expression for the
eigenvalue,  we return to the standard series solution and seek an asymptotic expansion. We
substitute the approximate solutions $S_1$ through $S_5$ into the stability matrix and compute analytically the eigenvalue using \textsc{Mathematica}; the resulting
eigenvalues are listed in Table~\ref{tab:eig_outphase}.

\begin{table}[htb]
\renewcommand{\arraystretch}{2.3}
$$\begin{array}{|c|c|}
\hline
S_n &  \text{Eigenvalue}\\
\hline
S_1 & \mathbbm{i}\sqrt{\e }\left( 2+\dfrac{45}{8}\e +\dfrac{375}{256}\e ^{2}+\dfrac{23253}{4096}\e ^{3}+\dfrac{5045483}{262144}\e^{4}\right)\\[0.5cm]
\hline
S_2 & \mathbbm{i}\sqrt{\e }\left( 2+4\e -\dfrac{135}{16}\e^{2}+\dfrac{657}{128}\e ^{3}+\dfrac{8311}{1024}\e^{4}\right)\\[0.5cm]
\hline
S_3 & \mathbbm{i}\sqrt{\e }\left( 2+4\e-5\e^{2}+\dfrac{3517}{128}\e ^{3}+\dfrac{131}{64}\e^{4}\right)\\[0.5cm]
\hline
S_4 & \mathbbm{i}\sqrt{\e }\left( 2+4\e-5\e^{2}+18\e ^{3}-\dfrac{15245}{256}\e^{4}\right)\\[0.5cm]
\hline
S_5 & \mathbbm{i}\sqrt{\e }\left( 2+4\e-5\e^{2}+18\e ^{3}-\dfrac{121}{4}\e^{4}\right)\\[0.5cm]
\hline
\end{array}$$
    \caption{The isolated eigenvalue of the out-of-phase configuration, obtained from the solution
    approximations $S_1$ through $S_5$.}
    \label{tab:eig_outphase}
\end{table}

\begin{figure}
\includegraphics[width=0.45\textwidth]{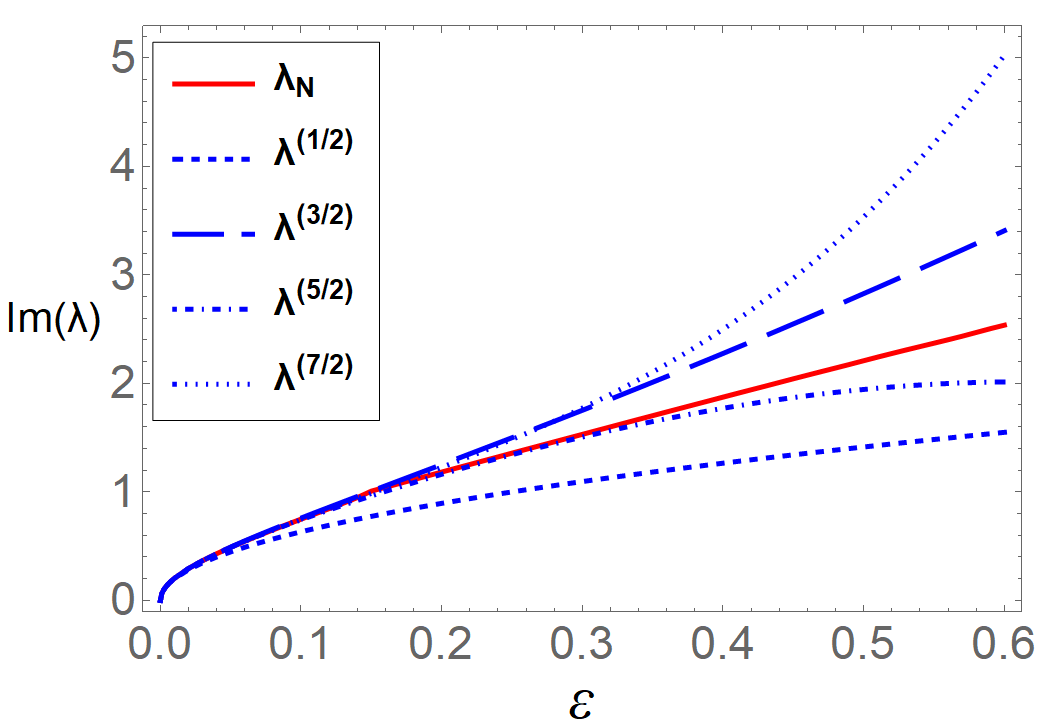}
\caption{Comparison between the numerical eigenvalue $\lambda_N$ (red,solid) and the asymptotic approximations $\lam{1/2}$ (blue, dashed),
$\lam{3/2}$ (blue, long-dashed), $\lam{5/2}$ (blue, dash-dotted) and $\lam{7/2}$ (blue, dotted) defined in the text, for the out-of-phase configuration.
The $\lam{5/2}$ approximation is the closest to the numerical result.}
\label{out_of_phase_li}
\end{figure}

Here too the coefficients of the eigenvalue expansion converge as the degree of the solution
increases. From the table we obtain the ${\cal O}(\e^{1/2})$, ${\cal O}(\e^{3/2})$,
${\cal O}(\e^{5/2})$ and ${\cal O}(\e^{7/2})$ truncated approximations of the imaginary part of the
isolated eigenvalue, namely $\lam{1/2}=2\e^{1/2}$, $\lam{3/2}=2\e^{1/2}+4\e ^{3/2}$,
$\lam{5/2}=2\e^{1/2}+4\e ^{3/2}-5\e ^{5/2}$ and
$\lam{7/2}=2\e^{1/2}+4\e ^{3/2}-5\e ^{5/2}+18\e ^{7/2}$, respectively. These are shown together with
the numerical eigenvalue $\lambda_N$ in Fig.~\ref{out_of_phase_li}; $\lam{1/2}$ is the estimate
reported in Ref.~\cite{PelinovskyKevrekidisFrantzeskakis2005a}. The $\lam{3/2}$ approximation
already captures the main behaviour of the numerical eigenvalue, and the inclusion of the next term
in $\lam{5/2}$ leads to a noticeably better agreement. Incorporating the subsequent $\e^{7/2}$
contribution in $\lam{7/2}$, on the other hand, does not improve matters; the resulting curve
diverges from the numerical eigenvalue faster than any of the others. As in the in-phase case, this
progressive deterioration is the signature of an asymptotic expansion, and $\lam{5/2}$ is again the
optimal truncation.

\begin{figure}
\includegraphics[width=0.45\textwidth]{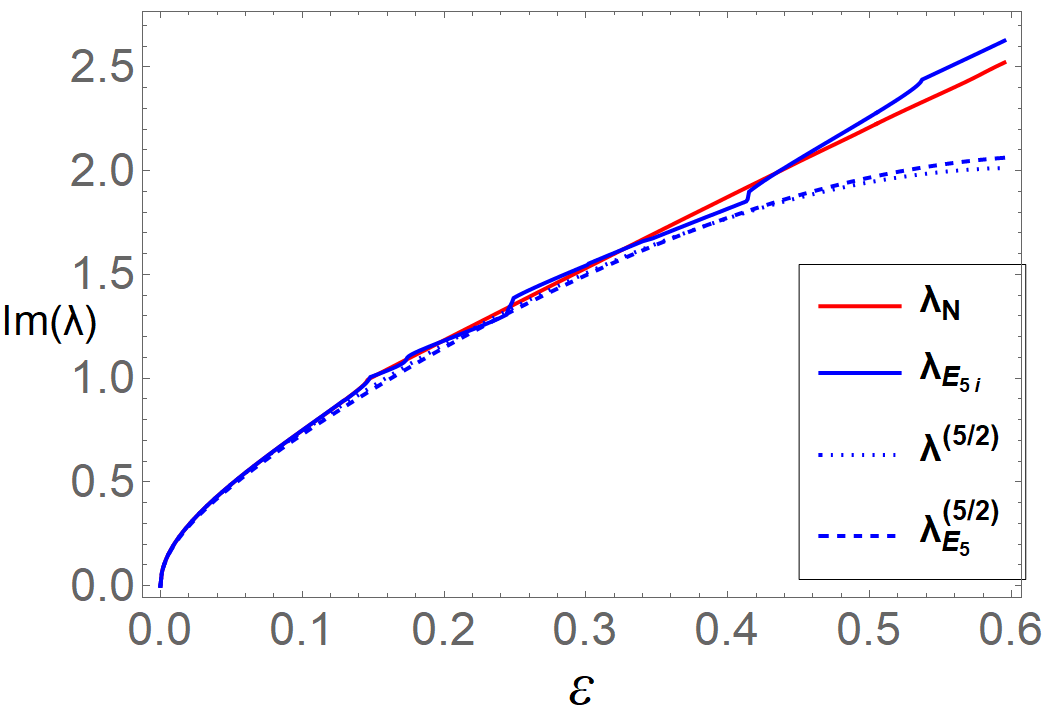}
\includegraphics[width=0.45\textwidth]{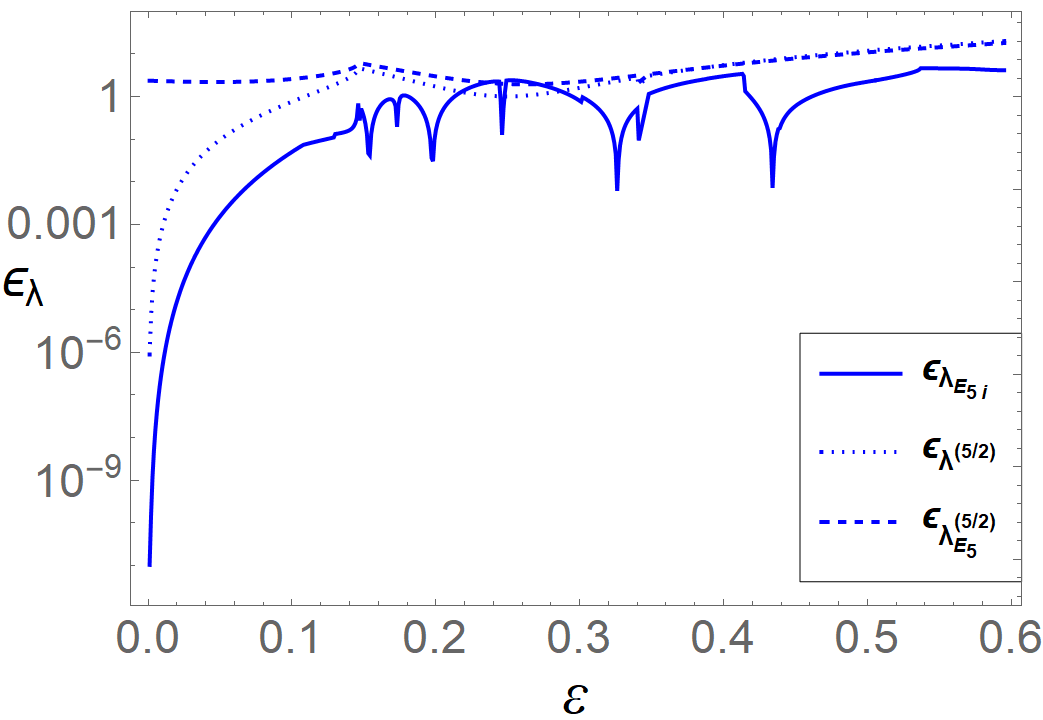}
\caption{Comparison of the positive imaginary part of the isolated eigenvalue of the out-of-phase
configuration, as obtained from the numerical solution $\lambda_N$ (red,solid), from the improved Euler transform $\lambda_{E_{5i}}$ (blue,solid), from the
series expansion truncated at the $\e^{5/2}$ term $\lambda^{(5/2)}$ (blue, dotted), and from the standard Euler transform $\lambda_{E_5}^{(5/2)}$ (blue,dashed) (left
panel). The corresponding relative errors with respect to the numerical eigenvalue are shown in the
right panel.}
\label{out_of_phase_sigrisi}
\end{figure}

Finally, applying the standard Euler transform $E_5$ to the $S_5$ solution and inserting the
transformed profile into the stability matrix gives, up to ${\cal O}(\e^{5/2})$,
$$\lambda^{5/2}_{E_{5}}=\dfrac{i\sqrt{\e }}{\sqrt{61}}\left( \dfrac{61}{4}+\dfrac{32063}{1024} \e - \dfrac{1174093393}{31981568} \e ^{2}\right).$$
This approximation is shown together with $\lambda_{E_{5i}}$, $\lam{5/2}$ and $\lambda_N$ in
Fig.~\ref{out_of_phase_sigrisi}, whose right panel again displays the corresponding relative errors
with respect to the numerical eigenvalue on a semilogarithmic scale.
Once again, $\lam{5/2}$ provides a reasonable approximation over a wide range
of $\varepsilon$ values.

\section{Conclusions and future directions}\label{sec:conclusions}
In this paper we constructed approximate discrete-soliton solutions of the cubic discrete nonlinear
Schr\"{o}dinger equation, for which exact analytical solutions are not available.
Yet, the model is one of wide physical applicability in systems
ranging from arrays of optical waveguides~\cite{AblowitzPrinariTrubatch,Kevrekidis2009} to
optical lattices of atomic condensates (with the latter also becoming
more experimentally accessible recently~\cite{haller}).
Three
configurations were considered: the standard single-site one and the two two-site ones, in-phase and
out-of-phase. The standard perturbation method delivers sign-alternating series expansions, which
makes the Euler transform, in both its standard and its improved form, a natural tool for the acceleration of the convergence of the solution. Both
transformations improved the accuracy of the solution appreciably and, importantly, helped to
preserve its exponentially localized character over a substantially wider range of couplings. It is
worth emphasizing how the improved transformation interpolates automatically between the plain
series at small $\e$ and the standard Euler transform at large $\e$, selecting in each case the
better one among the two. If such fine tuning is unnecessary, the standard transform provides adequate accuracy with a simpler, more standard algorithm.

As a benchmark of our approach we then characterized the stability of these solutions through the
eigenvalues of the stability matrix. Since the single-site configuration is spectrally stable
throughout the range considered, and therefore of no interest in this respect, we concentrated on
the two-site configurations. For the in-phase configuration we focused on the real, positive
(unstable) eigenvalue. A semi-numerical approach, in which the standard and improved-Euler profiles are evaluated
and inserted into the stability matrix, reproduces the fully numerical eigenvalue to better than
$1\%$ up to $\e=0.7$. Going further, we asked whether our approach could improve on the
leading-order estimate $2\sqrt{\e}$ of
Ref.~\cite{PelinovskyKevrekidisFrantzeskakis2005a}. Examining the successive truncations, we
concluded that the ${\cal O}(\e^{5/2})$ result derived from the standard series solution,
$\lam{5/2}=2\e^{1/2}-\e^{5/2}$, agrees best with the numerical eigenvalue and constitutes a
significant improvement on the previously available eigenvalue estimate; here the Euler transformation offers
no further gain. In addition, we also examined  the isolated eigenvalue of the out-of-phase configuration, as an
additional test of the method. The results parallel those of the in-phase case, with the added
richness --- and the added difficulty for the semi-numerical approach --- caused by the collisions
of the isolated eigenvalue with the continuous spectrum, which on the truncated lattice occur repeatedly
rather than once. Despite the limitations (e.g., as concerns the
real part of the approximation), the methodology provides a reasonable
(and far more widely applicable than previously available) estimate of the next order
correction approximation of the solutions' eigenvalues.

Overall, the Euler transform provides reasonably accurate analytical approximations of
discrete-soliton solutions while the use of the series solutions acquired by standard perturbation theory improves the analytical stability estimates available so far. Finally,
the semi-numerical variant of the approach reproduces the eigenvalue of interest to remarkable
accuracy. Two limitations should be kept in mind: the accuracy attainable is ultimately bounded by
the optimal truncation of an asymptotic expansion, and the number of non-zero sites retained by a
fifth-order expansion is what generates the spurious eigenvalue collisions observed in the
out-of-phase case. One can mitigate this issue by considering a larger
number of non-zero sites, at the potential additional expense of more cumbersome 
expressions.

Although the linear Euler transformation already yields significant improvements, 
a natural extension of this study would be to consider nonlinear resummation techniques such as the Pad\'{e} and Shanks
transformations~\cite{Pade1892, Shanks1955, BakerGravesMorris1996, Housiadas20171}, which are not restricted to
sign-alternating series and which we expect to improve the results further and to provide additional
insight into the system under consideration. Furthermore, while
in the present study we restricted consideration to one-dimensional
configurations, a natural question arises as to the applicability of
such techniques to more elaborate (yet still experimentally attainable~\cite{PRL1,PRL2}) configurations such as discrete vortices
in two-dimensional settings~\cite{PelinovskyKevrekidisFrantzeskakis2005b}
and perhaps even their more exotic extensions in three spatial dimensions~\cite{LUKAS2008339}. Such studies are currently in progress
and will be reported in future publications.

\bibliographystyle{apsrev4-2}
\bibliography{references}

\begin{thebibliography}{41}%
\makeatletter
\providecommand \@ifxundefined [1]{%
 \@ifx{#1\undefined}
}%
\providecommand \@ifnum [1]{%
 \ifnum #1\expandafter \@firstoftwo
 \else \expandafter \@secondoftwo
 \fi
}%
\providecommand \@ifx [1]{%
 \ifx #1\expandafter \@firstoftwo
 \else \expandafter \@secondoftwo
 \fi
}%
\providecommand \natexlab [1]{#1}%
\providecommand \enquote  [1]{``#1''}%
\providecommand \bibnamefont  [1]{#1}%
\providecommand \bibfnamefont [1]{#1}%
\providecommand \citenamefont [1]{#1}%
\providecommand \href@noop [0]{\@secondoftwo}%
\providecommand \href [0]{\begingroup \@sanitize@url \@href}%
\providecommand \@href[1]{\@@startlink{#1}\@@href}%
\providecommand \@@href[1]{\endgroup#1\@@endlink}%
\providecommand \@sanitize@url [0]{\catcode `\\12\catcode `\$12\catcode
  `\&12\catcode `\#12\catcode `\^12\catcode `\_12\catcode `\%12\relax}%
\providecommand \@@startlink[1]{}%
\providecommand \@@endlink[0]{}%
\providecommand \url  [0]{\begingroup\@sanitize@url \@url }%
\providecommand \@url [1]{\endgroup\@href {#1}{\urlprefix }}%
\providecommand \urlprefix  [0]{URL }%
\providecommand \Eprint [0]{\href }%
\providecommand \doibase [0]{https://doi.org/}%
\providecommand \selectlanguage [0]{\@gobble}%
\providecommand \bibinfo  [0]{\@secondoftwo}%
\providecommand \bibfield  [0]{\@secondoftwo}%
\providecommand \translation [1]{[#1]}%
\providecommand \BibitemOpen [0]{}%
\providecommand \bibitemStop [0]{}%
\providecommand \bibitemNoStop [0]{.\EOS\space}%
\providecommand \EOS [0]{\spacefactor3000\relax}%
\providecommand \BibitemShut  [1]{\csname bibitem#1\endcsname}%
\let\auto@bib@innerbib\@empty
\bibitem [{\citenamefont {Ablowitz}(2011)}]{ablowitz2011nonlinear}%
  \BibitemOpen
  \bibfield  {author} {\bibinfo {author} {\bibfnamefont {M.~J.}\ \bibnamefont
  {Ablowitz}},\ }\href {https://doi.org/10.1017/CBO9780511998324} {\emph
  {\bibinfo {title} {Nonlinear Dispersive Waves: Asymptotic Analysis and
  Solitons}}}\ (\bibinfo  {publisher} {Cambridge University Press},\ \bibinfo
  {year} {2011})\BibitemShut {NoStop}%
\bibitem [{\citenamefont {Scott}(2003)}]{Scott2003}%
  \BibitemOpen
  \bibfield  {author} {\bibinfo {author} {\bibfnamefont {A.}~\bibnamefont
  {Scott}},\ }\href@noop {} {\emph {\bibinfo {title} {Nonlinear Science:
  Emergence and Dynamics of Coherent Structures}}},\ \bibinfo {edition} {2nd}\
  ed.\ (\bibinfo  {publisher} {Oxford University Press},\ \bibinfo {address}
  {Oxford},\ \bibinfo {year} {2003})\BibitemShut {NoStop}%
\bibitem [{\citenamefont {Ablowitz}\ \emph {et~al.}(2004)\citenamefont
  {Ablowitz}, \citenamefont {Prinari},\ and\ \citenamefont
  {Trubatch}}]{AblowitzPrinariTrubatch}%
  \BibitemOpen
  \bibfield  {author} {\bibinfo {author} {\bibfnamefont {M.~J.}\ \bibnamefont
  {Ablowitz}}, \bibinfo {author} {\bibfnamefont {B.}~\bibnamefont {Prinari}},\
  and\ \bibinfo {author} {\bibfnamefont {A.~D.}\ \bibnamefont {Trubatch}},\
  }\href@noop {} {\emph {\bibinfo {title} {Discrete and Continuous Nonlinear
  {S}chr\"odinger Systems}}}\ (\bibinfo  {publisher} {Cambridge University
  Press, Cambridge},\ \bibinfo {year} {2004})\BibitemShut {NoStop}%
\bibitem [{\citenamefont {Dauxois}\ and\ \citenamefont
  {Peyrard}(2006)}]{Dauxois2006}%
  \BibitemOpen
  \bibfield  {author} {\bibinfo {author} {\bibfnamefont {T.}~\bibnamefont
  {Dauxois}}\ and\ \bibinfo {author} {\bibfnamefont {M.}~\bibnamefont
  {Peyrard}},\ }\href@noop {} {\emph {\bibinfo {title} {Physics of Solitons}}}\
  (\bibinfo  {publisher} {Cambridge University Press},\ \bibinfo {address}
  {Cambridge},\ \bibinfo {year} {2006})\BibitemShut {NoStop}%
\bibitem [{\citenamefont {Kevrekidis}(2009)}]{Kevrekidis2009}%
  \BibitemOpen
  \bibfield  {author} {\bibinfo {author} {\bibfnamefont {P.~G.}\ \bibnamefont
  {Kevrekidis}},\ }\href {https://doi.org/10.1007/978-3-540-89199-4} {\emph
  {\bibinfo {title} {The Discrete Nonlinear {S}chr\"{o}dinger Equation:
  Mathematical Analysis, Numerical Computations and Physical Perspectives}}},\
  \bibinfo {series} {Springer Tracts in Modern Physics}, Vol.\ \bibinfo
  {volume} {232}\ (\bibinfo  {publisher} {Springer},\ \bibinfo {address}
  {Berlin, Heidelberg},\ \bibinfo {year} {2009})\BibitemShut {NoStop}%
\bibitem [{\citenamefont {Kevrekidis}\ \emph {et~al.}(2015)\citenamefont
  {Kevrekidis}, \citenamefont {Frantzeskakis},\ and\ \citenamefont
  {Carretero-Gonz\'{a}lez}}]{KevrekidisBEC2015}%
  \BibitemOpen
  \bibfield  {author} {\bibinfo {author} {\bibfnamefont {P.~G.}\ \bibnamefont
  {Kevrekidis}}, \bibinfo {author} {\bibfnamefont {D.~J.}\ \bibnamefont
  {Frantzeskakis}},\ and\ \bibinfo {author} {\bibfnamefont {R.}~\bibnamefont
  {Carretero-Gonz\'{a}lez}},\ }\href@noop {} {\emph {\bibinfo {title} {The
  Defocusing Nonlinear {S}chr\"{o}dinger Equation: From Dark Solitons to
  Vortices and Vortex Rings}}}\ (\bibinfo  {publisher} {SIAM},\ \bibinfo
  {address} {Philadelphia},\ \bibinfo {year} {2015})\BibitemShut {NoStop}%
\bibitem [{\citenamefont {MacKay}\ and\ \citenamefont
  {Aubry}(1994)}]{MacKayAubry1994}%
  \BibitemOpen
  \bibfield  {author} {\bibinfo {author} {\bibfnamefont {R.~S.}\ \bibnamefont
  {MacKay}}\ and\ \bibinfo {author} {\bibfnamefont {S.}~\bibnamefont {Aubry}},\
  }\href {https://doi.org/10.1088/0951-7715/7/6/006} {\bibfield  {journal}
  {\bibinfo  {journal} {Nonlinearity}\ }\textbf {\bibinfo {volume} {7}},\
  \bibinfo {pages} {1623} (\bibinfo {year} {1994})}\BibitemShut {NoStop}%
\bibitem [{\citenamefont {Flach}\ and\ \citenamefont
  {Willis}(1998)}]{FlachWillis1998}%
  \BibitemOpen
  \bibfield  {author} {\bibinfo {author} {\bibfnamefont {S.}~\bibnamefont
  {Flach}}\ and\ \bibinfo {author} {\bibfnamefont {C.~R.}\ \bibnamefont
  {Willis}},\ }\href {https://doi.org/10.1016/S0370-1573(97)00068-9} {\bibfield
   {journal} {\bibinfo  {journal} {Physics Reports}\ }\textbf {\bibinfo
  {volume} {295}},\ \bibinfo {pages} {181} (\bibinfo {year}
  {1998})}\BibitemShut {NoStop}%
\bibitem [{\citenamefont {Flach}\ and\ \citenamefont
  {Gorbach}(2008)}]{FlachGorbach2008}%
  \BibitemOpen
  \bibfield  {author} {\bibinfo {author} {\bibfnamefont {S.}~\bibnamefont
  {Flach}}\ and\ \bibinfo {author} {\bibfnamefont {A.~V.}\ \bibnamefont
  {Gorbach}},\ }\href {https://doi.org/10.1016/j.physrep.2008.05.002}
  {\bibfield  {journal} {\bibinfo  {journal} {Physics Reports}\ }\textbf
  {\bibinfo {volume} {467}},\ \bibinfo {pages} {1} (\bibinfo {year}
  {2008})}\BibitemShut {NoStop}%
\bibitem [{\citenamefont {Aubry}(1997)}]{Aubry1997}%
  \BibitemOpen
  \bibfield  {author} {\bibinfo {author} {\bibfnamefont {S.}~\bibnamefont
  {Aubry}},\ }\href {https://doi.org/10.1016/S0167-2789(96)00261-8} {\bibfield
  {journal} {\bibinfo  {journal} {Physica D}\ }\textbf {\bibinfo {volume}
  {103}},\ \bibinfo {pages} {201} (\bibinfo {year} {1997})}\BibitemShut
  {NoStop}%
\bibitem [{\citenamefont {Aubry}(2006)}]{Aubry2006}%
  \BibitemOpen
  \bibfield  {author} {\bibinfo {author} {\bibfnamefont {S.}~\bibnamefont
  {Aubry}},\ }\href {https://doi.org/10.1016/j.physd.2005.12.020} {\bibfield
  {journal} {\bibinfo  {journal} {Physica D}\ }\textbf {\bibinfo {volume}
  {216}},\ \bibinfo {pages} {1} (\bibinfo {year} {2006})}\BibitemShut {NoStop}%
\bibitem [{\citenamefont {Pelinovsky}\ \emph
  {et~al.}(2005{\natexlab{a}})\citenamefont {Pelinovsky}, \citenamefont
  {Kevrekidis},\ and\ \citenamefont
  {Frantzeskakis}}]{PelinovskyKevrekidisFrantzeskakis2005a}%
  \BibitemOpen
  \bibfield  {author} {\bibinfo {author} {\bibfnamefont {D.~E.}\ \bibnamefont
  {Pelinovsky}}, \bibinfo {author} {\bibfnamefont {P.~G.}\ \bibnamefont
  {Kevrekidis}},\ and\ \bibinfo {author} {\bibfnamefont {D.~J.}\ \bibnamefont
  {Frantzeskakis}},\ }\href {https://doi.org/10.1016/j.physd.2005.07.021}
  {\bibfield  {journal} {\bibinfo  {journal} {Physica D}\ }\textbf {\bibinfo
  {volume} {212}},\ \bibinfo {pages} {1} (\bibinfo {year}
  {2005}{\natexlab{a}})}\BibitemShut {NoStop}%
\bibitem [{\citenamefont {Pelinovsky}\ \emph
  {et~al.}(2005{\natexlab{b}})\citenamefont {Pelinovsky}, \citenamefont
  {Kevrekidis},\ and\ \citenamefont
  {Frantzeskakis}}]{PelinovskyKevrekidisFrantzeskakis2005b}%
  \BibitemOpen
  \bibfield  {author} {\bibinfo {author} {\bibfnamefont {D.~E.}\ \bibnamefont
  {Pelinovsky}}, \bibinfo {author} {\bibfnamefont {P.~G.}\ \bibnamefont
  {Kevrekidis}},\ and\ \bibinfo {author} {\bibfnamefont {D.~J.}\ \bibnamefont
  {Frantzeskakis}},\ }\href {https://doi.org/10.1016/j.physd.2005.09.015}
  {\bibfield  {journal} {\bibinfo  {journal} {Physica D}\ }\textbf {\bibinfo
  {volume} {212}},\ \bibinfo {pages} {20} (\bibinfo {year}
  {2005}{\natexlab{b}})}\BibitemShut {NoStop}%
\bibitem [{\citenamefont {Lukas}\ \emph {et~al.}(2008)\citenamefont {Lukas},
  \citenamefont {Pelinovsky},\ and\ \citenamefont {Kevrekidis}}]{LUKAS2008339}%
  \BibitemOpen
  \bibfield  {author} {\bibinfo {author} {\bibfnamefont {M.}~\bibnamefont
  {Lukas}}, \bibinfo {author} {\bibfnamefont {D.}~\bibnamefont {Pelinovsky}},\
  and\ \bibinfo {author} {\bibfnamefont {P.}~\bibnamefont {Kevrekidis}},\
  }\href {https://doi.org/https://doi.org/10.1016/j.physd.2007.09.005}
  {\bibfield  {journal} {\bibinfo  {journal} {Physica D: Nonlinear Phenomena}\
  }\textbf {\bibinfo {volume} {237}},\ \bibinfo {pages} {339} (\bibinfo {year}
  {2008})}\BibitemShut {NoStop}%
\bibitem [{\citenamefont {Pelinovsky}\ and\ \citenamefont
  {Sakovich}(2011)}]{PelinovskySakovich2011}%
  \BibitemOpen
  \bibfield  {author} {\bibinfo {author} {\bibfnamefont {D.~E.}\ \bibnamefont
  {Pelinovsky}}\ and\ \bibinfo {author} {\bibfnamefont {A.}~\bibnamefont
  {Sakovich}},\ }\href {https://doi.org/10.1016/j.physd.2010.09.007} {\bibfield
   {journal} {\bibinfo  {journal} {Physica D}\ }\textbf {\bibinfo {volume}
  {240}},\ \bibinfo {pages} {265} (\bibinfo {year} {2011})}\BibitemShut
  {NoStop}%
\bibitem [{\citenamefont {Archilla}\ \emph {et~al.}(2003)\citenamefont
  {Archilla}, \citenamefont {Cuevas}, \citenamefont {Sánchez-Rey},\ and\
  \citenamefont {Alvarez}}]{ARCHILLA2003235}%
  \BibitemOpen
  \bibfield  {author} {\bibinfo {author} {\bibfnamefont {J.}~\bibnamefont
  {Archilla}}, \bibinfo {author} {\bibfnamefont {J.}~\bibnamefont {Cuevas}},
  \bibinfo {author} {\bibfnamefont {B.}~\bibnamefont {Sánchez-Rey}},\ and\
  \bibinfo {author} {\bibfnamefont {A.}~\bibnamefont {Alvarez}},\ }\href
  {https://doi.org/https://doi.org/10.1016/S0167-2789(03)00064-2} {\bibfield
  {journal} {\bibinfo  {journal} {Physica D: Nonlinear Phenomena}\ }\textbf
  {\bibinfo {volume} {180}},\ \bibinfo {pages} {235} (\bibinfo {year}
  {2003})}\BibitemShut {NoStop}%
\bibitem [{\citenamefont {Koukouloyannis}\ and\ \citenamefont
  {Kevrekidis}(2009)}]{KoukouloyannisKevrekidis2009}%
  \BibitemOpen
  \bibfield  {author} {\bibinfo {author} {\bibfnamefont {V.}~\bibnamefont
  {Koukouloyannis}}\ and\ \bibinfo {author} {\bibfnamefont {P.~G.}\
  \bibnamefont {Kevrekidis}},\ }\href
  {https://doi.org/10.1088/0951-7715/22/9/011} {\bibfield  {journal} {\bibinfo
  {journal} {Nonlinearity}\ }\textbf {\bibinfo {volume} {22}},\ \bibinfo
  {pages} {2269} (\bibinfo {year} {2009})}\BibitemShut {NoStop}%
\bibitem [{\citenamefont {Ahn}\ \emph {et~al.}(2001)\citenamefont {Ahn},
  \citenamefont {MacKay},\ and\ \citenamefont
  {Sepulchre}}]{AhnMacKaySepulchre2001}%
  \BibitemOpen
  \bibfield  {author} {\bibinfo {author} {\bibfnamefont {T.}~\bibnamefont
  {Ahn}}, \bibinfo {author} {\bibfnamefont {R.~S.}\ \bibnamefont {MacKay}},\
  and\ \bibinfo {author} {\bibfnamefont {J.-A.}\ \bibnamefont {Sepulchre}},\
  }\href {https://doi.org/10.1023/A:1012958817589} {\bibfield  {journal}
  {\bibinfo  {journal} {Nonlinear Dynamics}\ }\textbf {\bibinfo {volume}
  {25}},\ \bibinfo {pages} {157} (\bibinfo {year} {2001})}\BibitemShut
  {NoStop}%
\bibitem [{\citenamefont {Koukouloyannis}\ and\ \citenamefont
  {MacKay}(2005)}]{Koukouloyannis20051021}%
  \BibitemOpen
  \bibfield  {author} {\bibinfo {author} {\bibfnamefont {V.}~\bibnamefont
  {Koukouloyannis}}\ and\ \bibinfo {author} {\bibfnamefont {R.~S.}\
  \bibnamefont {MacKay}},\ }\href {https://doi.org/10.1088/0305-4470/38/5/004}
  {\bibfield  {journal} {\bibinfo  {journal} {Journal of Physics A:
  Mathematical and General}\ }\textbf {\bibinfo {volume} {38}},\ \bibinfo
  {pages} {1021 – 1030} (\bibinfo {year} {2005})},\ \bibinfo {note} {cited
  by: 24}\BibitemShut {NoStop}%
\bibitem [{\citenamefont {Cuevas-Maraver}\ \emph {et~al.}(2011)\citenamefont
  {Cuevas-Maraver}, \citenamefont {Koukouloyannis}, \citenamefont
  {Kevrekidis},\ and\ \citenamefont {Archilla}}]{Cuevas2011Multibreather}%
  \BibitemOpen
  \bibfield  {author} {\bibinfo {author} {\bibfnamefont {J.}~\bibnamefont
  {Cuevas-Maraver}}, \bibinfo {author} {\bibfnamefont {V.}~\bibnamefont
  {Koukouloyannis}}, \bibinfo {author} {\bibfnamefont {P.~G.}\ \bibnamefont
  {Kevrekidis}},\ and\ \bibinfo {author} {\bibfnamefont {J.~F.~R.}\
  \bibnamefont {Archilla}},\ }\href {https://doi.org/10.1142/S0218127411029690}
  {\bibfield  {journal} {\bibinfo  {journal} {International Journal of
  Bifurcation and Chaos}\ }\textbf {\bibinfo {volume} {21}},\ \bibinfo {pages}
  {2161} (\bibinfo {year} {2011})}\BibitemShut {NoStop}%
\bibitem [{\citenamefont {Koukouloyannis}\ \emph {et~al.}(2013)\citenamefont
  {Koukouloyannis}, \citenamefont {Kevrekidis}, \citenamefont {Cuevas},\ and\
  \citenamefont {Rothos}}]{KoukouloyannisKevrekidisCuevasRothos2013}%
  \BibitemOpen
  \bibfield  {author} {\bibinfo {author} {\bibfnamefont {V.}~\bibnamefont
  {Koukouloyannis}}, \bibinfo {author} {\bibfnamefont {P.~G.}\ \bibnamefont
  {Kevrekidis}}, \bibinfo {author} {\bibfnamefont {J.}~\bibnamefont {Cuevas}},\
  and\ \bibinfo {author} {\bibfnamefont {V.}~\bibnamefont {Rothos}},\ }\href
  {https://doi.org/10.1016/j.physd.2012.08.011} {\bibfield  {journal} {\bibinfo
   {journal} {Physica D}\ }\textbf {\bibinfo {volume} {242}},\ \bibinfo {pages}
  {16} (\bibinfo {year} {2013})}\BibitemShut {NoStop}%
\bibitem [{\citenamefont {Pelinovsky}\ and\ \citenamefont
  {Sakovich}(2012)}]{PelinovskySakovich2012}%
  \BibitemOpen
  \bibfield  {author} {\bibinfo {author} {\bibfnamefont {D.~E.}\ \bibnamefont
  {Pelinovsky}}\ and\ \bibinfo {author} {\bibfnamefont {A.}~\bibnamefont
  {Sakovich}},\ }\href {https://doi.org/10.1088/0951-7715/25/12/3423}
  {\bibfield  {journal} {\bibinfo  {journal} {Nonlinearity}\ }\textbf {\bibinfo
  {volume} {25}},\ \bibinfo {pages} {3423} (\bibinfo {year}
  {2012})}\BibitemShut {NoStop}%
\bibitem [{\citenamefont {Cuevas}\ \emph {et~al.}(2008)\citenamefont {Cuevas},
  \citenamefont {James}, \citenamefont {Kevrekidis}, \citenamefont {Malomed},\
  and\ \citenamefont
  {S\'{a}nchez-Rey}}]{CuevasJamesKevrekidisMalomedSanchezRey2008}%
  \BibitemOpen
  \bibfield  {author} {\bibinfo {author} {\bibfnamefont {J.}~\bibnamefont
  {Cuevas}}, \bibinfo {author} {\bibfnamefont {G.}~\bibnamefont {James}},
  \bibinfo {author} {\bibfnamefont {P.~G.}\ \bibnamefont {Kevrekidis}},
  \bibinfo {author} {\bibfnamefont {B.~A.}\ \bibnamefont {Malomed}},\ and\
  \bibinfo {author} {\bibfnamefont {B.}~\bibnamefont {S\'{a}nchez-Rey}},\
  }\href {https://doi.org/10.2991/jnmp.2008.15.s3.13} {\bibfield  {journal}
  {\bibinfo  {journal} {Journal of Nonlinear Mathematical Physics}\ }\textbf
  {\bibinfo {volume} {15}},\ \bibinfo {pages} {124} (\bibinfo {year}
  {2008})}\BibitemShut {NoStop}%
\bibitem [{\citenamefont {Bender}\ and\ \citenamefont
  {Orszag}(1999)}]{BenderOrszag1999}%
  \BibitemOpen
  \bibfield  {author} {\bibinfo {author} {\bibfnamefont {C.~M.}\ \bibnamefont
  {Bender}}\ and\ \bibinfo {author} {\bibfnamefont {S.~A.}\ \bibnamefont
  {Orszag}},\ }\href {https://doi.org/10.1007/978-1-4757-3069-2} {\emph
  {\bibinfo {title} {Advanced Mathematical Methods for Scientists and Engineers
  {I}: Asymptotic Methods and Perturbation Theory}}}\ (\bibinfo  {publisher}
  {Springer},\ \bibinfo {address} {New York},\ \bibinfo {year}
  {1999})\BibitemShut {NoStop}%
\bibitem [{\citenamefont {Hardy}(1949)}]{Hardy1949}%
  \BibitemOpen
  \bibfield  {author} {\bibinfo {author} {\bibfnamefont {G.~H.}\ \bibnamefont
  {Hardy}},\ }\href@noop {} {\emph {\bibinfo {title} {Divergent Series}}}\
  (\bibinfo  {publisher} {Oxford University Press},\ \bibinfo {address}
  {Oxford},\ \bibinfo {year} {1949})\BibitemShut {NoStop}%
\bibitem [{\citenamefont {Borel}(1899)}]{Borel1899}%
  \BibitemOpen
  \bibfield  {author} {\bibinfo {author} {\bibfnamefont {{\'E}.}~\bibnamefont
  {Borel}},\ }\href {https://doi.org/10.24033/asens.463} {\bibfield  {journal}
  {\bibinfo  {journal} {Annales scientifiques de l'\'{E}cole Normale
  Sup\'{e}rieure}\ }\textbf {\bibinfo {volume} {16}},\ \bibinfo {pages} {9}
  (\bibinfo {year} {1899})}\BibitemShut {NoStop}%
\bibitem [{\citenamefont {Pad\'{e}}(1892)}]{Pade1892}%
  \BibitemOpen
  \bibfield  {author} {\bibinfo {author} {\bibfnamefont {H.}~\bibnamefont
  {Pad\'{e}}},\ }\href {https://doi.org/10.24033/asens.378} {\bibfield
  {journal} {\bibinfo  {journal} {Annales scientifiques de l'\'{E}cole Normale
  Sup\'{e}rieure}\ }\textbf {\bibinfo {volume} {9}},\ \bibinfo {pages} {3}
  (\bibinfo {year} {1892})}\BibitemShut {NoStop}%
\bibitem [{\citenamefont {Shanks}(1955)}]{Shanks1955}%
  \BibitemOpen
  \bibfield  {author} {\bibinfo {author} {\bibfnamefont {D.}~\bibnamefont
  {Shanks}},\ }\href {https://doi.org/10.1002/sapm19553411} {\bibfield
  {journal} {\bibinfo  {journal} {Journal of Mathematics and Physics}\ }\textbf
  {\bibinfo {volume} {34}},\ \bibinfo {pages} {1} (\bibinfo {year}
  {1955})}\BibitemShut {NoStop}%
\bibitem [{\citenamefont {Wynn}(1956)}]{Wynn1956}%
  \BibitemOpen
  \bibfield  {author} {\bibinfo {author} {\bibfnamefont {P.}~\bibnamefont
  {Wynn}},\ }\href {https://doi.org/10.2307/2002183} {\bibfield  {journal}
  {\bibinfo  {journal} {Mathematical Tables and Other Aids to Computation}\
  }\textbf {\bibinfo {volume} {10}},\ \bibinfo {pages} {91} (\bibinfo {year}
  {1956})}\BibitemShut {NoStop}%
\bibitem [{\citenamefont {{van Wijngaarden}}(1965)}]{VanWijngaarden1965}%
  \BibitemOpen
  \bibfield  {author} {\bibinfo {author} {\bibfnamefont {A.}~\bibnamefont {{van
  Wijngaarden}}},\ }in\ \href@noop {} {\emph {\bibinfo {booktitle} {Cursus:
  Wetenschappelijk rekenen {B}: Onderdeel ``{P}roces {A}nalyse''}}}\ (\bibinfo
  {publisher} {Stichting Mathematisch Centrum},\ \bibinfo {address}
  {Amsterdam},\ \bibinfo {year} {1965})\BibitemShut {NoStop}%
\bibitem [{\citenamefont {Levin}(1973)}]{Levin1973}%
  \BibitemOpen
  \bibfield  {author} {\bibinfo {author} {\bibfnamefont {D.}~\bibnamefont
  {Levin}},\ }\href {https://doi.org/10.1080/00207167308803075} {\bibfield
  {journal} {\bibinfo  {journal} {International Journal of Computer
  Mathematics}\ }\textbf {\bibinfo {volume} {3}},\ \bibinfo {pages} {371}
  (\bibinfo {year} {1973})}\BibitemShut {NoStop}%
\bibitem [{\citenamefont {Weniger}(1989)}]{Weniger1989}%
  \BibitemOpen
  \bibfield  {author} {\bibinfo {author} {\bibfnamefont {E.~J.}\ \bibnamefont
  {Weniger}},\ }\href {https://doi.org/10.1016/0167-7977(89)90011-7} {\bibfield
   {journal} {\bibinfo  {journal} {Computer Physics Reports}\ }\textbf
  {\bibinfo {volume} {10}},\ \bibinfo {pages} {189} (\bibinfo {year}
  {1989})}\BibitemShut {NoStop}%
\bibitem [{\citenamefont {Baker}\ and\ \citenamefont
  {Graves-Morris}(1996)}]{BakerGravesMorris1996}%
  \BibitemOpen
  \bibfield  {author} {\bibinfo {author} {\bibfnamefont {G.~A.}\ \bibnamefont
  {Baker}, \bibfnamefont {Jr.}}\ and\ \bibinfo {author} {\bibfnamefont
  {P.}~\bibnamefont {Graves-Morris}},\ }\href@noop {} {\emph {\bibinfo {title}
  {{P}ad\'{e} Approximants}}},\ \bibinfo {edition} {2nd}\ ed.,\ \bibinfo
  {series} {Encyclopedia of Mathematics and its Applications}, Vol.~\bibinfo
  {volume} {59}\ (\bibinfo  {publisher} {Cambridge University Press},\ \bibinfo
  {address} {Cambridge},\ \bibinfo {year} {1996})\BibitemShut {NoStop}%
\bibitem [{\citenamefont {Brezinski}\ and\ \citenamefont
  {Redivo~Zaglia}(1991)}]{BrezinskiRedivoZaglia1991}%
  \BibitemOpen
  \bibfield  {author} {\bibinfo {author} {\bibfnamefont {C.}~\bibnamefont
  {Brezinski}}\ and\ \bibinfo {author} {\bibfnamefont {M.}~\bibnamefont
  {Redivo~Zaglia}},\ }\href@noop {} {\emph {\bibinfo {title} {Extrapolation
  Methods: Theory and Practice}}},\ \bibinfo {series} {Studies in Computational
  Mathematics}, Vol.~\bibinfo {volume} {2}\ (\bibinfo  {publisher}
  {North-Holland},\ \bibinfo {address} {Amsterdam},\ \bibinfo {year}
  {1991})\BibitemShut {NoStop}%
\bibitem [{\citenamefont {Sidi}(2003)}]{Sidi2003}%
  \BibitemOpen
  \bibfield  {author} {\bibinfo {author} {\bibfnamefont {A.}~\bibnamefont
  {Sidi}},\ }\href@noop {} {\emph {\bibinfo {title} {Practical Extrapolation
  Methods: Theory and Applications}}},\ \bibinfo {series} {Cambridge Monographs
  on Applied and Computational Mathematics}, Vol.~\bibinfo {volume} {10}\
  (\bibinfo  {publisher} {Cambridge University Press},\ \bibinfo {address}
  {Cambridge},\ \bibinfo {year} {2003})\BibitemShut {NoStop}%
\bibitem [{\citenamefont {Housiadas}(2017)}]{Housiadas20171}%
  \BibitemOpen
  \bibfield  {author} {\bibinfo {author} {\bibfnamefont {K.~D.}\ \bibnamefont
  {Housiadas}},\ }\href {https://doi.org/10.1016/j.jnnfm.2017.05.005}
  {\bibfield  {journal} {\bibinfo  {journal} {Journal of Non-Newtonian Fluid
  Mechanics}\ }\textbf {\bibinfo {volume} {247}},\ \bibinfo {pages} {1 – 14}
  (\bibinfo {year} {2017})}\BibitemShut {NoStop}%
\bibitem [{\citenamefont {Abramowitz}\ and\ \citenamefont
  {Stegun}(1965)}]{Euler_initial}%
  \BibitemOpen
  \bibinfo {editor} {\bibfnamefont {M.}~\bibnamefont {Abramowitz}}\ and\
  \bibinfo {editor} {\bibfnamefont {I.~A.}\ \bibnamefont {Stegun}},\ eds.,\
  \href@noop {} {\emph {\bibinfo {title} {Handbook of mathematical functions
  with formulas, graphs, and mathematical tables}}}\ (\bibinfo  {publisher}
  {Dover Publications},\ \bibinfo {year} {1965})\BibitemShut {NoStop}%
\bibitem [{\citenamefont {van Wijngaarden}(1965)}]{Euler_transform}%
  \BibitemOpen
  \bibfield  {author} {\bibinfo {author} {\bibfnamefont {A.}~\bibnamefont {van
  Wijngaarden}},\ }\href@noop {} {\emph {\bibinfo {title} {Proces Analyse}}},\
  \bibinfo {type} {Cursus: Wetenschappelijk Rekenen B}\ (\bibinfo
  {institution} {Stichting Mathematisch Centrum},\ \bibinfo {address}
  {Amsterdam},\ \bibinfo {year} {1965})\ \bibinfo {note} {mathematisch Centrum,
  Amsterdam}\BibitemShut {NoStop}%
\bibitem [{\citenamefont {Cruickshank}\ \emph {et~al.}(2025)\citenamefont
  {Cruickshank}, \citenamefont {Lorenzi}, \citenamefont {La~Rooij},
  \citenamefont {Kerr}, \citenamefont {Hilker}, \citenamefont {Kuhr},
  \citenamefont {Salasnich},\ and\ \citenamefont {Haller}}]{haller}%
  \BibitemOpen
  \bibfield  {author} {\bibinfo {author} {\bibfnamefont {R.}~\bibnamefont
  {Cruickshank}}, \bibinfo {author} {\bibfnamefont {F.}~\bibnamefont
  {Lorenzi}}, \bibinfo {author} {\bibfnamefont {A.}~\bibnamefont {La~Rooij}},
  \bibinfo {author} {\bibfnamefont {E.~F.}\ \bibnamefont {Kerr}}, \bibinfo
  {author} {\bibfnamefont {T.}~\bibnamefont {Hilker}}, \bibinfo {author}
  {\bibfnamefont {S.}~\bibnamefont {Kuhr}}, \bibinfo {author} {\bibfnamefont
  {L.}~\bibnamefont {Salasnich}},\ and\ \bibinfo {author} {\bibfnamefont
  {E.}~\bibnamefont {Haller}},\ }\href {https://doi.org/10.1103/sh72-wnmv}
  {\bibfield  {journal} {\bibinfo  {journal} {Phys. Rev. Lett.}\ }\textbf
  {\bibinfo {volume} {135}},\ \bibinfo {pages} {263404} (\bibinfo {year}
  {2025})}\BibitemShut {NoStop}%
\bibitem [{\citenamefont {Neshev}\ \emph {et~al.}(2004)\citenamefont {Neshev},
  \citenamefont {Alexander}, \citenamefont {Ostrovskaya}, \citenamefont
  {Kivshar}, \citenamefont {Martin}, \citenamefont {Makasyuk},\ and\
  \citenamefont {Chen}}]{PRL1}%
  \BibitemOpen
  \bibfield  {author} {\bibinfo {author} {\bibfnamefont {D.~N.}\ \bibnamefont
  {Neshev}}, \bibinfo {author} {\bibfnamefont {T.~J.}\ \bibnamefont
  {Alexander}}, \bibinfo {author} {\bibfnamefont {E.~A.}\ \bibnamefont
  {Ostrovskaya}}, \bibinfo {author} {\bibfnamefont {Y.~S.}\ \bibnamefont
  {Kivshar}}, \bibinfo {author} {\bibfnamefont {H.}~\bibnamefont {Martin}},
  \bibinfo {author} {\bibfnamefont {I.}~\bibnamefont {Makasyuk}},\ and\
  \bibinfo {author} {\bibfnamefont {Z.}~\bibnamefont {Chen}},\ }\href
  {https://doi.org/10.1103/PhysRevLett.92.123903} {\bibfield  {journal}
  {\bibinfo  {journal} {Phys. Rev. Lett.}\ }\textbf {\bibinfo {volume} {92}},\
  \bibinfo {pages} {123903} (\bibinfo {year} {2004})}\BibitemShut {NoStop}%
\bibitem [{\citenamefont {Fleischer}\ \emph {et~al.}(2004)\citenamefont
  {Fleischer}, \citenamefont {Bartal}, \citenamefont {Cohen}, \citenamefont
  {Manela}, \citenamefont {Segev}, \citenamefont {Hudock},\ and\ \citenamefont
  {Christodoulides}}]{PRL2}%
  \BibitemOpen
  \bibfield  {author} {\bibinfo {author} {\bibfnamefont {J.~W.}\ \bibnamefont
  {Fleischer}}, \bibinfo {author} {\bibfnamefont {G.}~\bibnamefont {Bartal}},
  \bibinfo {author} {\bibfnamefont {O.}~\bibnamefont {Cohen}}, \bibinfo
  {author} {\bibfnamefont {O.}~\bibnamefont {Manela}}, \bibinfo {author}
  {\bibfnamefont {M.}~\bibnamefont {Segev}}, \bibinfo {author} {\bibfnamefont
  {J.}~\bibnamefont {Hudock}},\ and\ \bibinfo {author} {\bibfnamefont {D.~N.}\
  \bibnamefont {Christodoulides}},\ }\href
  {https://doi.org/10.1103/PhysRevLett.92.123904} {\bibfield  {journal}
  {\bibinfo  {journal} {Phys. Rev. Lett.}\ }\textbf {\bibinfo {volume} {92}},\
  \bibinfo {pages} {123904} (\bibinfo {year} {2004})}\BibitemShut {NoStop}%
\bibitem [{\citenamefont {Lederer}\ \emph {et~al.}(2008)\citenamefont
  {Lederer}, \citenamefont {Stegeman}, \citenamefont {Christodoulides},
  \citenamefont {Assanto}, \citenamefont {Segev},\ and\ \citenamefont
  {Silberberg}}]{LEDERER20081}%
  \BibitemOpen
  \bibfield  {author} {\bibinfo {author} {\bibfnamefont {F.}~\bibnamefont
  {Lederer}}, \bibinfo {author} {\bibfnamefont {G.~I.}\ \bibnamefont
  {Stegeman}}, \bibinfo {author} {\bibfnamefont {D.~N.}\ \bibnamefont
  {Christodoulides}}, \bibinfo {author} {\bibfnamefont {G.}~\bibnamefont
  {Assanto}}, \bibinfo {author} {\bibfnamefont {M.}~\bibnamefont {Segev}},\
  and\ \bibinfo {author} {\bibfnamefont {Y.}~\bibnamefont {Silberberg}},\
  }\href {https://doi.org/10.1016/j.physrep.2008.04.004} {\bibfield  {journal}
  {\bibinfo  {journal} {Phys. Rep.}\ }\textbf {\bibinfo {volume} {463}},\
  \bibinfo {pages} {1} (\bibinfo {year} {2008})}\BibitemShut {NoStop}%
\bibitem [{\citenamefont {Morsch}\ and\ \citenamefont
  {Oberthaler}(2006)}]{morsch}%
  \BibitemOpen
  \bibfield  {author} {\bibinfo {author} {\bibfnamefont {O.}~\bibnamefont
  {Morsch}}\ and\ \bibinfo {author} {\bibfnamefont {M.}~\bibnamefont
  {Oberthaler}},\ }\href {https://doi.org/10.1103/RevModPhys.78.179} {\bibfield
  {journal} {\bibinfo  {journal} {Rev. Mod. Phys.}\ }\textbf {\bibinfo
  {volume} {78}},\ \bibinfo {pages} {179} (\bibinfo {year}
  {2006})}\BibitemShut {NoStop}%
\bibitem [{\citenamefont {Vuli{\'c}}\ \emph {et~al.}(2026)\citenamefont
  {Vuli{\'c}}, \citenamefont {{\v{S}}anti{\'c}}, \citenamefont {Buljan},\ and\
  \citenamefont {Aumiler}}]{Vulic2026}%
  \BibitemOpen
  \bibfield  {author} {\bibinfo {author} {\bibfnamefont {V.}~\bibnamefont
  {Vuli{\'c}}}, \bibinfo {author} {\bibfnamefont {N.}~\bibnamefont
  {{\v{S}}anti{\'c}}}, \bibinfo {author} {\bibfnamefont {H.}~\bibnamefont
  {Buljan}},\ and\ \bibinfo {author} {\bibfnamefont {D.}~\bibnamefont
  {Aumiler}},\ }\href {https://doi.org/10.1002/lpor.71994} {\bibfield
  {journal} {\bibinfo  {journal} {Laser Photonics Rev.}\ }\bibinfo {pages}
  {e71994} (\bibinfo {year} {2026})}\BibitemShut {NoStop}%
\end{thebibliography}%

\appendix
\section{Euler transform: single-site configuration}
\label{EulerSingle}
In this appendix we present the full transformed solution for the single-site configuration,
obtained with the improved Euler transformation according to the algorithm of
Sec.~\ref{sec:Euler}.

Throughout the appendix, ${\cal S}_n$ denotes the result of the standard perturbation method for a
single site of the lattice, the subscript $n$ being the number of terms retained. By ${\cal E}_n$ we
denote the truncated series obtained after applying the standard Euler transform to ${\cal S}_n$,
and by ${\cal E}_{ni}^m$ the improved Euler transform acting on ${\cal S}_n$, with the subscript
``$i$'' standing for ``improved'' and the superscript $m$ labelling which of the $n$ possible
series, according to the conditions of Table~\ref{tab:Eulers}, is selected for the given value of
$\e$.

These tables confirm what was observed in Fig.~\ref{on_site_error}: the improved Euler algorithm
always selects the series with the smaller error. For small $\e$ the improved transform
${\cal E}_{ni}^n$ coincides with the original series ${\cal S}_n$, whereas for large $\e$ it becomes
the standard Euler transform, ${\cal E}_{ni}^1={\cal E}_n$. Note that the branch intervals are in
general unions of disjoint sub-intervals, i.e.\ the algorithm switches back and forth between
branches as $\e$ increases; this is the origin of the spikes observed in
Figs.~\ref{in_phase_sigrisi}, \ref{out_euler_improved} and~\ref{out_of_phase_sigrisi}.
The interval endpoints below are quoted to two decimal places because they were obtained on a grid of $\Delta\varepsilon = 0.01$ and they are not the exact endpoints which correspond to the satisfaction of the conditions of section~\ref{sec:Euler}. 

\begin{table}[!ht]
\renewcommand{\arraystretch}{1.75}
\begin{tabular}{c|c|cl}
Oscillator &$\e$& & \qquad\qquad\qquad Transformed series\\\hline
$\vi_{0}$\quad$ $&
$[0, 0.09]$ &$ $\quad $ $&
${\cal S}_{5}={\cal E}_{5i}^{5}=1+\varepsilon-\dfrac{3\varepsilon ^{2}}{2}+\dfrac{7\varepsilon ^{3}}{2}-\dfrac{85\varepsilon ^{4}}{8}+\dfrac{287 \varepsilon ^{5}}{8} $\\[0.3cm]
&$[0.1, 0.11]\cup [0.28,0.31]$ & &
${\cal E}_{5i}^{4}=1+\varepsilon-\dfrac{3\varepsilon ^{2}}{2}+\dfrac{7\varepsilon ^{3}}{2}-\dfrac{255\varepsilon ^{4}}{32}+\dfrac{287 \varepsilon ^{5}}{32} $\\[0.3cm]
&$ [0.12, 0.27] \cup [0.32,0.37]$ & &
${\cal E}_{5i}^{3}=1+\varepsilon-\dfrac{3\varepsilon ^{2}}{2}+\dfrac{49\varepsilon ^{3}}{16}-\dfrac{85\varepsilon ^{4}}{16}+\dfrac{287 \varepsilon ^{5}}{64} $\\[0.3cm]
&$[0.38, 0.53]$ & &
${\cal E}_{5i}^{2}=1+\varepsilon-\dfrac{45\varepsilon ^{2}}{32}+\dfrac{77\varepsilon ^{3}}{32}-\dfrac{425\varepsilon ^{4}}{128}+\dfrac{287 \varepsilon ^{5}}{128} $\\[0.3cm]
&$[0.54, 1]$ & &
${\cal E}_{5}={\cal E}_{5i}^{1}=1+\dfrac{31\varepsilon}{32}-\dfrac{39\varepsilon ^{2}}{32}+\dfrac{7\varepsilon ^{3}}{4}-\dfrac{255\varepsilon ^{4}}{128}+\dfrac{287 \varepsilon ^{5}}{256} $\\[0.3cm]\hline
$\vi_{1}$\quad$ $&
$[0, 0.1]$ &$ $\quad $ $&
${\cal S}_{5}={\cal E}_{5i}^{5}=\varepsilon -\varepsilon ^{2}+\dfrac{5\varepsilon ^{3}}{2}-\dfrac{15\varepsilon ^{4}}{2}+\dfrac{195 \varepsilon ^{5}}{8} $\\[0.3cm]
&$\lbrace 0.11\rbrace \cup [0.3,0.32]$ & &
${\cal E}_{5i}^{4}=\varepsilon -\varepsilon ^{2}+\dfrac{5\varepsilon ^{3}}{2}-\dfrac{45\varepsilon ^{4}}{8}+\dfrac{195 \varepsilon ^{5}}{32} $\\[0.3cm]
&$[0.12, 0.29] \cup [0.33,0.36]$ & &
${\cal E}_{5i}^{3}=\varepsilon -\varepsilon ^{2}+\dfrac{35\varepsilon ^{3}}{16}-\dfrac{15\varepsilon ^{4}}{4}+\dfrac{195 \varepsilon ^{5}}{64} $\\[0.3cm]
&$[0.37, 0.63]$ & &
${\cal E}_{5i}^{2}=\varepsilon -\dfrac{15\varepsilon ^{2}}{16}+\dfrac{55\varepsilon ^{3}}{32}-\dfrac{75\varepsilon ^{4}}{32}+\dfrac{195 \varepsilon ^{5}}{128} $\\[0.3cm]
&$[0.64, 1]$ & &
${\cal E}_{5}={\cal E}_{5i}^{1}=\dfrac{31\varepsilon}{32}-\dfrac{13\varepsilon ^{2}}{16}+\dfrac{5\varepsilon ^{3}}{4}-\dfrac{45\varepsilon ^{4}}{32}+\dfrac{195 \varepsilon ^{5}}{256} $ \\[0.3cm]\hline
$\vi_{2}$\quad$ $&
$[0, 0.1]$ &$ $\quad $ $&
${\cal S}_{4}={\cal E}_{4i}^{4}=\varepsilon^{2}- 3\varepsilon ^{3}+\dfrac{19\varepsilon ^{4}}{2}-\dfrac{63\varepsilon ^{5}}{2} $\\[0.3cm]
&$\left\lbrace 0.11 \right\rbrace \cup [0.31,0.32]$ & &
${\cal E}_{4i}^{2}=\varepsilon ^{2}-\dfrac{21\varepsilon ^{3}}{8}+\dfrac{19\varepsilon ^{4}}{4}-\dfrac{63 \varepsilon ^{5}}{16} $\\[0.3cm]
&$ [0.12, 0.3] \cup [0.33,1]$ & &
${\cal E}_{4}={\cal E}_{4i}^{1}=\dfrac{15\varepsilon ^{2}}{16}-\dfrac{33\varepsilon ^{3}}{16}+\dfrac{95\varepsilon ^{4}}{32}-\dfrac{63 \varepsilon ^{5}}{32} $\\[0.3cm]\hline
$\vi_{3}$\quad$ $&
$[0, 0.06]$ &$ $\quad $ $&
${\cal S}_{3}={\cal E}_{3i}^{3}=\varepsilon^{3}- 5\varepsilon ^{4}+\dfrac{41\varepsilon ^{5}}{2} $\\[0.3cm]
&$ [0.23, 0.26]$ & &
${\cal E}_{3i}^{2}=\varepsilon^{3}- \dfrac{15\varepsilon ^{4}}{4}+\dfrac{41\varepsilon ^{5}}{8} $\\[0.3cm]
&$[0.07, 0.22]\cup [0.27,1]$ & &
${\cal E}_{3}={\cal E}_{3i}^{1}=\dfrac{7\varepsilon^{3}}{8}- \dfrac{5\varepsilon ^{4}}{2}+\dfrac{41\varepsilon ^{5}}{16} $\\[0.3cm]\hline
$\vi_{4}$\quad$ $&
$[0, 0.04]$ &$ $\quad $ $&
${\cal S}_{2}={\cal E}_{2i}^{2}=\varepsilon^{4}-7\varepsilon ^{5} $\\[0.3cm]
&$[0.05, 1]$ & &
${\cal E}_{2}={\cal E}_{2i}^{1}= \dfrac{3\varepsilon ^{4}}{4}-\dfrac{7\varepsilon ^{5}}{4} $\\[0.3cm]\hline
$\vi_{5}$\quad$ $&
$[0, 1]$ &$ $\quad $ $&
${\cal S}_{1}={\cal E}_{1}={\cal E}_{1i}^{1}=\varepsilon^{5} $\\[0.2cm]\hline
\end{tabular}
\caption{Transformed series for the single-site configuration, listed by lattice site and by the
range of $\e$ over which each branch of the improved Euler transformation is selected.}
\label{tab:app_single}
\end{table}

\end{document}